\documentclass{llncs}

\usepackage{amsfonts}
\usepackage{yfonts}  
\usepackage{upgreek}
\usepackage{amssymb}
\usepackage{textgreek}
\usepackage{colortbl}
\usepackage{amsmath}
\usepackage{enumerate}
\usepackage{appendix}

\begin{document}
\newcommand{\li}{\underline}
\newcommand{\PP}{{\cal P}}
\newcommand{\RR}{{\cal R}}
\newcommand{\lam}{\lambda}
\newcommand{\ei}{\epsilon_i}
\newcommand{\be}{\bar{\epsilon}}
\newcommand{\eproof}{\hfill $\Box$ \vspace{0.2cm}}
\newtheorem{rem}{Remark}
\newtheorem{lemm}{Lemma}
\newtheorem{exam}{Example}
\newtheorem{theor}{Theorem}
\newtheorem{prop}{Proposition}
\newtheorem{corol}{Corollary}
\newtheorem{defi}{Definition}
\newtheorem{notat}{Notation}
\newcommand{\conjug}[1]{{\fontdimen8\textfont3=0.25pt\mkern2mu\overline{\mkern-1mu #1\mkern-2mu}}\mkern2mu}

\title{\bf Post-quantum encryption algorithms of high-degree 3-variable polynomial congruences: \\
BS cryptosystems and BS key generation}

\author{Nicholas J. Daras\inst{}}
\institute{Hellenic Military Academy, Department of Mathematics and Engineering Sciences, 16673 Vari Attikis, Greece.
    \email{Email:njdaras@sse.gr}}

\date{}
\maketitle\thispagestyle{plain} \pagestyle{plain}
\vspace{-\baselineskip}

\begin{abstract} \noindent\textit{We will construct post-quantum encryption algorithms based on three-variable polynomial Beal-Schur congruence. After giving a proof of Beal's conjecture and citing some applications of it to selected cases where the discrete logarithm and some of its generalizations are unsolvable problems,  we will investigate the formulation and validity of an appropriate version of the Beal's conjecture on finite fields of integers. In contrast to the infinite case, we will show that  the corresponding Beal-Schur congruence equation $x^{p}+y^{q}\equiv z^{r} (mod \mathcal{N})$ has non-trivial solutions into the finite field $\mathbb{Z}_{\mathcal{N}} $, for all sufficiently large primes $\mathcal{N}$ that do not divide the product $xyz$, under certain mutual divisibility conditions of the exponents $p$, $q$ and $r$. We will apply this result to generate the so-called BS cryptosystems, i.e., simple and secure post-quantum encryption algorithms based on the Beal-Schur congruence equation, as well as  new cryptographic key generation methods, whose post-quantum algorithmic encryption security relies on having an infinite number of options for the parameters $p$, $q$, $r$, $\mathcal{N}$.}  
\end{abstract}

\section{Introduction}

\indent Encryption using Diophantine equations is important because it provides a high level of security and efficiency in post-quantum encryption systems. Diophantine equations are used to generate spoofing and spreading keys, which are suitable for data encryption. Using chaotic systems to generate the parameters of the Diophantine equations adds a layer of security to the encryption process ([22, 27]). By solving Diophantine equations, it is possible to change the position of bits and pixels in the encryption sequence, which can increase the security of the ciphertext [22]. This approach can also generate a large number of hash keys and their distributions, improving the security of the encryption scheme ([27]). In general, encryption using Diophantine equations provides a secure and efficient way to protect sensitive data.\\

\indent In this direction, many notable papers have been written on the subject of the implementation of post-quantum encryption methods based on the solution of various Diophantine equations. Without being able to include all of them, let me mention a select few, without implying that my list is exhaustive. The interested reader is referred to these articles, as well as to the works mentioned in their relevant bibliography.\\
\indent Claude Shannon seems to have been the first to realize that cryptosystems containing Diophantine difficulties have the greatest uncertainty in the key selection process ([46]). \\
\indent Several years later, in 1995, Chia-Hung Lin, Chen Chung Chang, and Richard Chia-Tung Lee, respecting this pioneering view of Shannon, proposed a new public key (two-key) encryption scheme ([33]). In their encryption scheme, keys can be easily generated and encryption and decryption methods are easily generated. To encrypt a message, the sender must perform a cross-product of the message being sent and the enciphering key. The receiver can easily decrypt it by conducting several multiplication operations and modulus operations. For the security analysis, the authors also investigated some possible attacks on the presented scheme.  \\
\indent Three years later, in 1998, M. S. Ibrahim, F. S. Abd Rabou and H. Y. Zorkta introduced an algorithm for data security within computer networks ([28]). Their algorithm was based on a public key cipher system and Diophantine equations. To reduce the possibility of message length prediction, the algorithm was improved by using dummy bits’ technique. In addition, some mathematical rules are used to limit the length of the encryption key. For the problem of truncation in floating point type representation of numbers, a proposed solution was introduced for obtaining a very long integer type representation.  \\
\indent After the lapse of some years, in 2014, Attila Bérczes, Lajos Hajdu, Noriko Hirata-Kohno, Tünde Kovàcs and Attila Pethö presented a cryptosystem with a new key exchange protocol based on Diophantine equations of polynomial type ([6]). Their protocol was inspired by that of H. Yosh whose security was based on the interpretation of Diophantine equations ([48]). They proposed a key exchange protocol relying on the hardness of solving Diophantine equations in the ring of S-integers. Researching post-quantum cryptography is now an important task. Although various candidates of post-quantum cryptosystems have been constructed, sizes of their public keys are large.  \\
\indent One year later, in 2015 September, Valeriy O.Osipyan developed the mathematical model of a post-quantum cryptosystem based on the gamma superposition method, in which the algorithm of the inverse transformation of the closed text is reduced to the impossibility of problem solution ([44]). He extended the multiplicative knapsack task and considered the problem of working out of alphabetic cryptosystems mathematical models. His research was based on earlier pioneer ideas by Claude Shannon, who believed, that \textit{cryptosystems containing Diophantine difficulties possess the greatest uncertainty of key selection process}. Necessary and sufficient conditions at which generalized multiplicative knapsack is injective on $\mathbb{Z}_{\mathcal{N}} $, $\mathcal{N}\geq 2$, were also established.  \\
\indent As a follow-up to these ideas, three years later, in 2018, Valeriy O. Osipyan and Kirill I. Litvinov presented a mathematical model of information security system based on the linear inhomogeneous Diophantine equation ([45]). According to this model, the plain text is the solution of the Diophantine equation and the cipher text is the right side of equation. They also provided a method to find this solution, based on the construction of a system of equations whose solution is the desired solution of the original Diophantine equation. The system of equations is created using some hidden information. \\
\indent In addition to all these valuable contributions, in 2015, Shinya Okumura proposed a new public key cryptosystem based on Diophantine equations which is called of degree increasing type ([43]). He used an analogous method to the \textit{Algebraic Surface Cryptosystem} proposed by Koichiro Akiyama, Yasuhiro Goto and Hideyuki Miyake ([1]) to construct a candidate of post-quantum cryptosystem whose security is expected to be based on certain Diophantine equations. Okumura’s analysis suggested that Diophantine equations achieve high security with small public key sizes. \\
\indent Just three years later, Jintai Ding, Momonari Kudo, Shinya Okumura, Tsuyo-Shi Takagi and Chengdong Tao proposed a polynomial time-attack on the one-way property of Diophantine equations ([21]). They reduced the security of Diophantine equations to finding special short lattice points of some low-rank lattices derived from public data. The usual LLL algorithm (:Lenstra–Lenstra–Lovász lattice basis reduction algorithm) could not find the most important lattice point in our experiments because of certain properties of the lattice point. Their heuristic analysis leads us to using a variant of the LLL algorithm, called the weighted LLL algorithm. However, their experiments suggested that Diophantine equations with 128-bit security becomes insecure by their attack. \\
\indent Two years before, in 2016, Shuhong Gao and Raymond Heindl, inspired by earlier work by Lih-Chung Wang, Bo-Yin Yang, Yuh-Hua Hu and Feipei Lai ([49]) had presented a general framework for Multivariate Public Key Cryptosystems, which combines ideas from both triangular and oil-vinegar schemes ([23]). Within this framework, they had proposed a new public key cryptosystem based on a solution of a Diophantine equation over polynomial rings. \\
\indent In the year 2022, Avinash Vijayarangan, Veena Narayanan, Vijayarangan Natarajan, and Srikanth Raghavendran elucidated some geometric properties of positive integral solutions of the quadratic Diophantine equation $x_{1}^{2} + x_{2}^{2}= y_{1}^{2}+y_{2}^{2}$ ($x_{1}$, $x_{2}$, $y_{1}$, $y_{2} > 0$), as well as their use in communication protocols ([35]). Given a pair $(x_{1}, y_{1})$, it was a challenge to find another pair $(x_{2}, y_{2})$ satisfying the quadratic Diophantine $x_{1}^{2} + x_{2}^{2}= y_{1}^{2}+y_{2}^{2}$.  \\
\indent From the point of view of ensuring security, a particularly promising evolution of cryptographic algorithms based on Diophantine equations is that which uses \textbf{\textit{polynomial congruences over finite fields of positive integers}}. \\
\indent In this context, perhaps the most representative example is the \textit{Multivariate Public-Key Cryptography}. Following Jintai Ding and Bo-Yin Yang ([20]) it is the study of public key cryptosystems where the trapdoor one-way function takes the form of a multivariate quadratic polynomial map over a finite field. Namely, the public key is in general given by a set of \textbf{quadratic} polynomials \\

\indent \hspace{3.5cm} $P = \left( p_{1} \left( w_{1},...,w_{n} \right) ,...,p_{m} \left( w_{1},...,w_{n} \right)  \right) $, \\
 
\noindent where each $p_{i}$ is a (usu. quadratic) nonlinear polynomial in $w = \left( w_{1},...,w_{n} \right)$:\\

\indent \hspace{1.5cm}$z_{k} = p_{k} (w):= \sum_{i} P_{i,k}w_{i} + \sum_{i} Q_{i,k}w_{i}^{2}+ \sum_{i, j} R_{i,j,k}w_{i}w_{j}$\\

\noindent with all coefficients and variables in $\mathbb{K} = \mathbb{F}_\mathcal{N}$, the field with $\mathcal{N}$ elements. The evaluation of these polynomials at any given value corresponds to either the encryption procedure or the verification procedure. Such public key cryptosystems are called multivariate public key cryptosystems. Inverting a multivariate quadratic map is equivalent to solving a set of quadratic equations over a finite field, or the following problem:\\

\indent \hspace{1cm} Problem $\mathcal{MQ}$: Solve the system \\
\indent \hspace{3cm}$p_{1}(x) = p_{2}(x) = ...= p_{m}(x) = 0$, \\
\indent \hspace{1cm}where each $p_{i}$ is a quadratic in $x = \left( x_{1},...,x_{n} \right) $. \\
\indent \hspace{1cm}All coefficients and variables are in $\mathbb{K} = \mathbb{F}_\mathcal{N}$, the field with $\mathcal{N}$ elements.\\

$\mathcal{MQ}$ is in general an NP-hard problem. Such problems are believed to be hard unless the class P is equal to NP. Of course, a random set of quadratic equations would not have a trapdoor and hence not be usable in a multivariate public key cryptosystem. The corresponding mathematical structure to a system of polynomial equations, not necessarily generic, is the ideal generated by those polynomials. So, philosophically speaking, multivariate cryptography relates to mathematics that handles polynomial ideals, namely algebraic geometry.\\
\indent However, since we are no longer dealing with “random” or “generic” systems, but systems where specific trapdoors exist, the security of multivariate public key cryptosystems is then not guaranteed by the NP-hardness of MQ, and effective attacks may exist for any chosen trapdoor. Section 5 of [20] describes the most important attack modes against multivariate public key cryptosystems (Linearization Equations, Lazard-Faugère System Solvers, Differential Attacks, Rank Attacks, Distilling Oil from Vinegar and Other Attacks on Unbalanced Oil-Vinegar).\\
\indent Undoubtedly, one of the basic tools we have for solving quadratic or linear Diophantine equations over finite fields of integers is the method of Coppersmith. Coppersmith’s method [14] uses lattice basis reduction to find small solutions of polynomial congruences. This method and its variants have been used to solve a number of problems across cryptography, including attacks against low public exponent RSA [13, 14], demonstrating the insecurity of small private exponent RSA, factoring with partial knowledge [13, 14], and the approximate integer common divisor problem.\\

\indent Let's list some basic literature on the Coppersmith method. In [14] and its references, Coppersmith considered monic polynomials $f(x)\in \mathbb{Z}[x]$ of degree $\delta$ and a modulus $\mathcal{N}$. He gave an algorithm that is polynomial time in $log N$ and $2^{\delta}$ for finding all integers $x$ such that \\

\indent \hspace{3.5cm} $f(x) \equiv 0 \left( mod \mathcal{N} \right)$ and $|x|< N^{1/\delta}$. \\

\noindent The strategy is to use the LLL algorithm to produce a polynomial $h(x)\in Q[x]$ such that $h(x) = 0$ for all integers $x$ with these properties. \\
\indent Coppersmith also considered irreducible two variable polynomials \\

\indent \hspace{3.5cm} $p(x,y) = \sum_{i,j} p_{i,j} x^{i} y^{j} \in \mathbb{Z}[x,y]$. \\

\noindent Suppose $X,Y > 0$ are given and that \\

\indent \hspace{4.5cm} $W = max_{i,j} \vert p_{i,j}\vert X^{i} Y^{j}$. \\

\noindent Coppersmith gave an algorithm for finding an integer solution $(x,y)$ of $p(x,y)=0$ such that $\vert x \vert \leq X $ and $\vert y \vert \leq Y$, if one exists, provided that $XY < W^{3/2 \delta}$. The approach was to generate a polynomial $h(x,y) \in \mathbb{Q}[x,y]$ not divisible by $p(x,y)$ such that $h(x,y) = 0$ for all integers $x$ and $y$ as above. Since $p(x,y)$ was assumed to be irreducible, the common zero locus of $h(x,y)$ and $p(x,y)$ is finite. \\
\indent Coppersmith also pointed out in [14] that the method for deriving auxiliary functions which must vanish at suitably small solutions of Diophantine equalities or congruences can be applied to more general systems of polynomial equations in several variables. The reader interested in a literature pertaining to this issue is referred to [35] and [40] and the references therein. Without trying to provide an overview of this literature, allow me to mention a few specific results. \\
\indent The paper [37] by 37.	Alexander May and Maike Ritzenhofen addresses the problem of finding all small integer solutions to systems of one variable polynomial congruences modulo a set of mutually co-prime moduli. The plan of action was to show this is equivalent to finding all small solutions of a different congruence involving only one polynomial in one variable. Following a remark by Coppersmith in [15], the authors noted that one cannot in general improve the upper bound $ |x|< \sigma^{1/\delta}$ to $|x|< \sigma^{1/\delta+\epsilon}$ for $ \epsilon>0 $ in Coppermith’s original theorem due to the fact that that the number of solutions may then become exponential in $\sigma$. This will certainly be the case when $f(x)= x^{\delta}$ and $\sigma=\mathcal{N}^{\delta}$ for some prime $\mathcal{N}$, for example. A central question has been to determine when the following heuristic holds. When the geometry of numbers applied to suitable convex symmetric sets of polynomials produces one auxiliary polynomial that must vanish on the desired solution, will there be sufficiently many auxiliary polynomials in a slightly larger convex set of this kind which are algebraically independent? The common zero locus of this larger set of polynomials will then be finite.\\

\indent The paper [2] by Aurélie Bauer and Antoine Joux begins with a discussion of various experiments to test the above heuristic. It then develops a sufficient condition for Coppersmith's method to succeed in finding three algebraically independent auxiliary polynomials when applied to finding small integral zeroes of a polynomial in three variables. This is based on Bauer's work [3], who constructed three variable examples in which the heuristic fails. Additional instances where the heuristic fails for bivariate polynomials over integers are discussed in [36], accompanied by a commentary of the experimental evidence, both in favour and against the heuristic. The discussion after Figure 3 of [36] of the situations in which the heuristic did not hold experimentally should be also noted.\\
\indent In [11] Ted Chinburg, Brett Hemenway, Nadia Heninger and Zachary Scherr showed that for one variable polynomial congruences of the kind Coppersmith originally considered, there are no auxiliary polynomials of the kind his method requires if one replaces the bound $|x|< \sigma^{1/\delta}$ by $|x|< \sigma^{1/\delta+\epsilon}$ for some $ \epsilon>0 $. This is because the capacity theory on curves discussed in the next section shows that there are infinitely many solutions of the same congruence in the ring of all algebraic integers whose conjugates all lie in the disk of radius $\sigma^{1/\delta+\epsilon}$. All such algebraic integers would have to be a zero of an auxiliary polynomial of the kind used in Coppersmith’s method. Theorem 1.4 of [12] shows how capacity theory can give a necessary and sufficient condition for the above heuristic to hold in the case of two variable linear polynomial congruences. In this case, the heuristic holds for a positive proportion of the possible initial data, and it fails for a different positive proportion of this data.\\
\indent In practice, all of these cryptographic schemes mentioned so far contain and are based on linear or quadratic polynomials that have small degrees. In contrast to all of these, the cryptographic schemes we will consider below are based on \textbf{polynomial congruence equations of \textit{$3$ variables}} \textbf{with high order exponents (: \textit{arbitrarily greater than $2$})}. Specifically,  in this article, new and simple post-quantum encryption schemes based on  Diophantine equations defined by three-variable polynomial congruences will be given, and,  particularly,  secure algorithmic encryption frameworks will established based on the Beal-Schur congruence equation  $x^{p}+y^{q}\equiv z^{r} (mod \mathcal{N})$ into a finite field  $\mathbb{Z}_{\mathcal{N}} $ where the integer $\mathcal{N}$ should be chosen to be a sufficiently large prime number and the exponents $p$, $q$ and $r$ can be chosen to be arbitrarily large. As will be proved later (Theorem 2), \textit{under certain general conditions, the equation in question has non trivial solutions $x$, $y$ and $z$ into a finite field} $\mathbb{Z}_{\mathcal{N}} $.
Although Bauer's promising method ([2] and [3]), which generalizes  Coppersmith's pioneer ideas, aspires to be able to solve such three-variable polynomial parity equations, at least on a theoretical level (since the computational complexity of the method remains unknown), all post-quantum schemes encryption schemes proposed here ensure total security, as any cryptanalysis attempt appears ``dead end", since the use of such an agreement equation implies the construction of systems that have a large variety of equally likely keys (in fact, there are an infinite number of key choices). And only one key is correct... \\

\indent At this point it should be made clear that any act to extend Beal's equation to the infinite ring $\mathbb{Z}$ of the integers (or to the monoid  $\mathbb{N}_{0}=\mathbb{N}\cup\left\lbrace 0 \right\rbrace $  of the natural numbers including $0$) is impossible as to the existence of non-trivial solutions. Indeed, according to Beal's famous conjecture, \textit{for any given natural numbers $p$, $q$ and $r$, there exists no non-trivial triplet of natural numbers $x$, $y$, $z$ that do not have a common prime divisor and solve the Diophantine equation } $x^{p}+y^{q}= z^{r} $. In the next Section 2, we will give a constructive proof of Beal's Conjecture ([16]). The proof will be given explicitly in Theorem 1 and requires a series of elementary techniques and auxiliary Propositions, all of which are included in Appendix A. Some subsequent propositions arising from the validity of Beal's conjecture will be proposed in Section 3. All these resulting applications concern mainly common cases in which it is impossible to solve the discrete logarithm problem, the double discrete logarithm problem ([38]) and the multidimensional discrete logarithm problem as well as its variants ([24, 41]). As it was already pointed out, contrary to the case of the ring of integers, in Section 4, we will show that the "local" version of Beal's conjecture does not hold proving that the corresponding congruence equation $x^{p}+y^{q}\equiv z^{r} (mod \mathcal{N}) $ has a non-trivial solution into the finite field $\mathbb{Z}_{\mathcal{N}} $ for all primes $\mathcal{N}$ that are enough large and such that $\mathcal{N}$ does not divide $xyz $ under some necessary conditions that the exponents $p$, $q$ and $r$ must satisfy (Theorem 2). The proof we will apply is directly inspired by the Fourier analytic approach adopted in [31, 52] in order to prove Schur's theorem ([47, 31]). As in the case of Theorem 1, the relevant proof will be given explicitly and requires a series of elementary techniques and auxiliary Propositions, all of which are included in Appendix B. Finally, in the concluding Sections 5 and 6, we will use the positive result of Theorem 2, in order to generate three simple and strong algorithmic encryption frameworks, as well as three corresponding cryptographic key generation frameworks, whose security (and thus the difficulty of breaking them) is based, on the one hand, in the existence of an infinite number of appropriate choices for the four parameters $p$, $q$, $r$ and $\mathcal{N}$ to solve the Beal-Schur congruence equation $x^{p}+y^{q}\equiv z^{r} (mod \mathcal{N}) $ in a finite field $\mathbb{Z}_{\mathcal{N}} $ (with the integer $\mathcal{N}$ to be, in any case, a sufficiently large prime) and, on the other hand, to the given impossibility of solving the equation, since, at least so far, no way of solving it has been found and, moreover, it does not seem likely to achieve a solution method at least for the next few years...

\section{A Constructive Proof of Beal's Conjecture}

In the early twentieth century, David Hilbert presented twenty-three great mathematical problems which featured main directions of scientific research throughout the period that followed ([25,26]). By analogy, in 2016, John F. Nash and Michael Th. Rassias gave a list with seventeen, currently unsolved problems in modern mathematics, in the belief that these problems are expected to determine several of the main research directions at least during the beginning of the XXI century ([42]). The third problem in the series of this list refers to the exploration of the possibility of extending Fermat's Last Theorem.\\
\indent Even before achieving a proof of this Theorem ([50]), various generalizations had already been considered, to equations of the shape $Ax^{p}+ By^{q}=Cz^{r}$, for fixed integers $A$,$B$ and $C$. In this direction, the Theorem of Henri Darmon and Andrew Granville states that \textit{if} $A$ \textit{,}$B$\textit{,} $C$\textit{,} $p$\textit{,} $q$ \textit{and} $r$ \textit{are fixed integers with} $p^{-1}+q^{-1}+r^{-1}<1$, \textit{the generalized Fermat equation} $Ax^{p}+ By^{q}=Cz^{r}$ \textit{has at most finitely many solutions in coprime non-zero integers} $x$,$y$ \textit{and} $z$ ([18]). However, as is made clear through those mentioned by Michael A. Bennett, Imin Chen, Sander R. Dahmen and Soroosh Yazdani ([5, 42]), except the solutions identified by Preda Mih{\u a}ilescu in the Catalan equation ([39]) and the solutions derived from some elementary numerical identities where at least one among the exponents $p$, $q$ and $r$ equals $2$ ([42], page 175), \textit{in all other known cases}, \textit{there is no non-trivial solution of this equation once we assume that} $A=B=C=1$ and $p,q,r\geq2$ ([4, 5, 6, 7, 10, 26, 30]). \\
\indent In this context, \textit{Beal's Conjecture} claims that \textit{if} $min\lbrace p,q,r\rbrace\geq 3$, \textit{there are no non-trivial coprime solutions} (: $gcd\left(  x,y,z\right) =1$)\textit{ to the generalized Fermat equation} $x^{p}+y^{q}=z^{r}$ ([17, 42]). To date, many computational attempts produced strong evidence that such a claim may be correct ([34]. The  aim of this Section is to give a constructive proof of Beal's Conjecture into the ring of integer numbers ([16]). \\

\indent To get a contradiction, let us suppose that, on the contrary, Beal's Conjecture does not apply. This is equivalent to assume that there are three integer exponents $p$, $q$, $r\geq 3$ and three coprime positive integers $x=even$, $y=odd$ and $z=odd$ satisfying a generalized Fermat equation of the form  $x^{p} + y^{q}=z^{r}$.\\

 \indent Given any $m$, $n$, $\lambda$ in $\mathbb{N}=\mathbb{Ζ}^{+}$ (:the set of all natural numbers, that is of all positive integers), $\lambda=odd$ and $\sigma\in \mathbb{Ζ}^{+} \setminus \left\lbrace 1 \right\rbrace $, put
\begin{center}
$\tau_{m,n}:=-y^{\sigma q-4}n(2m+n)+\tau_{0}$, with $\tau_{0}:=\lambda y^{q-2}$.
\end{center} 

\noindent Adopting this definition, it is clear that $\tau_{m,n}$ is an integer and the following result applies.

\begin{lemm}  
\textit{There exist} $m_{*}$, $n_{*}\in \normalfont \mathbb{Ζ}^{+}$, \textit{for which the corresponding expressions} $z^{2}- y^{2}\tau_{m_{*},n_{*}} $  \textit{represent prime numbers.}
\end{lemm}
\noindent \textit{Proof}. Since $gcd \left(  - y^{\sigma q-2},  \left[z^{2} - \tau_{0} y^{2}\right]\right) =1$ (\textit{Proposition 1} in \textit{Appendix Α}), an application of Dirichlet's Theorem on arithmetic progressions in its basic form, shows that there are many $m_{*} \in \mathbb{Ζ}^{+}$ and  $\nu \in \mathbb{Ζ}^{+}$ such that $y^{\sigma q-2} \left(  m_{*}^{2} -\nu^{2} \right)  - \left[z^{2}- \tau_{0} y^{2} \right] = \pi$ with $\pi$=prime number (see \textit{Proposition 2} in \textit{Appendix Α}). The integers $n_{*}=-m_{*} \pm \nu $ will be the two roots of the equation $y^{\sigma q - 2} n_{*}^{2}+ 2m_{*} y^{\sigma q-2} n_{*} + \left[ z^{2}- \tau_{0} y^{2} \right]- \pi =0$ (see \textit{Proposition 3} in \textit{Appendix Α}). We infer 
\[ y^{\sigma q-2} n_{*}^{2} + 2m_{*} y^{\sigma q-2} n_{*} + \left[ z^{2} - \tau_{0} y^{2}\right]-\pi = z^{2} -  y ^{2} \tau_{m_{*},n_{*}} - \pi=0\] 
(see \textit{Proposition 4} in \textit{Appendix Α}) and the proof is completed. $\square $\\

We may now draw the following  conclusion.

\begin{lemm} 
If $m_{*}$ and $n_{*}$ are as in Lemma 1 and  $\tau_{m_{*},n_{*}} z^{r-2} -y^{q-2} \neq \pm1$, then  
\begin{center}  
$ gcd \left( x^{p-2}, \tau_{m_{*},n_{*}} z^{r-2}-y^{q-2}\right)=1$.
\end{center} 
\end{lemm}
\noindent \textit{Proof}. To get a contradiction, suppose there is a natural number $ \varrho>1 $ such that $x^{p-2}=\varrho c_{1}$ and $\tau_{m_{*}, n_{*}}z^{r-2}-y^{q-2}=\varrho c_{3}$ for some integers $c_{1}$ and $c_{3}$. Multiplication of the first equation by $x^{2}$ gives \\
\indent \hspace{4cm} $ x^{p}=\varrho \left( c_{1} x^{2} \right) $,\\
\noindent while multiplication of the second one by $ y^{2}$ and $z^{2}$ gives
\begin{center}
$ y^{2}\tau_{m_{*}, n_{*}}z^{r-2}- y^{q}=  \varrho \left( c_{3} y^{2}\right) \Rightarrow  y^{q}= y^{2}\tau_{m_{*}, n_{*}}z^{r-2} - \varrho \left( c_{3} y^{2}\right)$ and
\end{center}
\begin{center}
$\tau_{m_{*}, n_{*}}z^{r}-z^{2}y^{q-2}=\varrho \left( c_{3}z^{2}\right) \Rightarrow \tau_{m_{*}, n_{*}}z^{r}=z^{2}y^{q-2}+ \varrho \left( c_{3} z^{2}\right)$,
\end{center} 
\noindent respectively. Adding the first two equations, we get $x^{p} + y^{q} = \varrho \left( c_{1} x^{2} - c_{3} y^{2}\right)$  $+ y^{2}\tau_{m_{*}, n_{*}}z^{r-2}$ or, by hypothesis, $ z^{r}=\varrho \left(c_{1} x^{2} - c_{3} y^{2}\right) + y^{2}\tau_{m_{*}, n_{*}}z^{r-2}$ which can equivalently be stated as
\begin{center}
$z^{r-2} \left[ z^{2}- y^{2} \tau_{m_{*}, n_{*}}\right]= \varrho \left( c_{1} x^{2} - c_{3} y^{2} \right) $.
\end{center}
Similarly, subtracting the third equation above from the multiple of the first one by $ \tau_{m_{*}, n_{*}}$, we get $\tau_{m_{*}, n_{*}}\left(x^{p}- z^{r} \right) = \varrho \left(\tau_{m_{*}, n_{*}} c_{1} x^{2}-c_{3} z^{2} \right) -z^{2}y^{q-2}$, or, by hypothesis, $-\tau_{m_{*}, n_{*}} y^{q}=  \varrho ( \tau_{m_{*}, n_{*}}c_{1} x^{2} -$ $c_{3} z^{2} ) - z^{2}y^{q-2}$ which can equivalently be written as
\begin{center}
$y^{q-2} \left[z^{2} - y^{2}\tau_{m_{*}, n_{*}} \right]=\varrho \left( \tau_{m_{*}, n_{*}}c_{1}x^{2} - c_{3} z^{2} \right) $.
\end{center}
\noindent Having regard to \textit{Lemma 1}, from the last two relationships, it follows that $\varrho$ divides both $z^{r-2}$ and $y^{q-2}$. This contradicts the requirement $ gcd \left(  x, y, z \right)=1$ under the assumption that $x^{p}+ y^{q}=z^{r}$. So, it is impossible to have $gcd \lbrace x^{p-2}, \tau_{m_{*}, n_{*}}$ $z^{r-2} - y^{q-2} \rbrace >1$, and the proof is completed. $\square $

\begin{rem} 
{\normalfont If} $p = 2${\normalfont , the proof of} Lemma 2 {\normalfont would be impossible to start, for the reason that our original assumption} $x^{p-2}=\varrho c_{1}${\normalfont would have degenerated into the false relationship} $1=\varrho c_{1}${\normalfont, with} $\varrho, c_{1}\in \mathbb{Z}$ {\normalfont and $ \varrho>1$.} $\square $
\end{rem}

\indent We are now in position to prove the main result of this paper.

\begin{theorem} 
If $ x^{p}+ y^{q}=z^{r}$, where $x$, $y$, $z$, $p$, $q$ and $r$ are positive integers and $p$, $q$ and $r$ are all greater than $2$, then $x$, $y$ and $z$ must have a common prime factor.
\end{theorem}

\noindent \textit{Proof}. We have to show that there is no non-trivial positive solution to the equation $ x^{p}+ y^{q}=z^{r}$, 
provided that $ min\left\lbrace p,q,r \right\rbrace >2$ and $gcd \left(  x, y, z \right)=1$. Let $m_{*}$ and $n_{*}$ be as in \textit{Lemma 1}, satisfying  $\tau_{m_{*},n_{*}} z^{r-2} -y^{q-2} \neq \pm1$. Let also $U, V, \tilde{U},\tilde{V}\in \mathbb{Z}\setminus\lbrace0\rbrace$ with $U\tilde{V} - \tilde{U}V\neq0 $. Then the system
\[\left( \mathbb{S}_{\mathbb{Ζ}}\right) 
\left\{\begin{array}{ll}
 \ (U)X+ (-V)Y+(\tau_{m_{*},n_{*}}V)Z= 1\\
 \ (\tilde{U})X+ (-\tilde{V})Y+(\tau_{m_{*},n_{*}}\tilde{V})Z= 1\\
 \ ( x^{2})X + ( y^{2})Y+(-z^{2})Z=0, 
\end{array}\right.
\]
with determinant $\left( U\tilde{V}-\tilde{U}V\right)\left( z^{2} -  y^{2}\tau_{m_{*},n_{*}}\right)\neq0$ (see \textit{Proposition 5} in \textit{Appendix Α}), has unique solution\\

$X=- \frac{V-\tilde{V}}{ U\tilde{V}-\tilde{U}V}$, \\

$Y=- \frac{\left( U-\tilde{U}\right) z^{2}+\left( V-\tilde{V}\right)\tau_{m_{*},n_{*}} x^{2}}{\left( U\tilde{V}-\tilde{U}V\right)\left( z^{2} -  y^{2}\tau_{m_{*},n_{*}}\right)}$ and \\

$Z= -\frac{\left( U-\tilde{U}\right) y^{2}\pm \left( V-\tilde{V}\right) x^{2}}{\left( U\tilde{V}-\tilde{U}V\right)\left( z^{2} -  y^{2}\tau_{m_{*},n_{*}}\right)}$ (see \textit{Proposition 6} in \textit{Appendix Α}).\\

\indent By \textit{Lemma 2}, we have $ gcd \left(  x^{p-2}, \tau_{m_{*},n_{*}} z^{r-2}-y^{q-2}\right)=1$, so an application of Bezout's identity guarantees that the Diophantine equation 
\begin{center}
$(x^{p-2})u+(\tau_{m_{*},n_{*}}z^{r-2} -y^{q-2})v= 1$ 
\end{center}
can be solved in $\mathbb{Ζ}^{2}$. So, given any fixed partial integer solution $(u, v)\in \mathbb{Ζ}^{2}$ of this equation and any $\kappa \in \mathbb{Ζ}^{+}$, $\ell \in \mathbb{Ζ}^{+}$, $\kappa\neq \ell $, we can take the induced solutions $U$, $V$, $\tilde{U}$, $\tilde{V}\in \mathbb{Ζ}\setminus \lbrace0\rbrace$ defined by\\

\indent $U:=u+\kappa\left( \tau_{m_{*},n_{*}} z^{r-2}-y^{q-2} \right)$, $V:=v-\kappa x^{p-2}$, \\
\indent $\tilde{U}:=u + \ell \left( \tau_{m_{*},n_{*}} z^{r-2}-y^{q-2} \right)$ and $\tilde{V}:=v-\ell x^{p-2}$.\\

\noindent Notice that, by construction, \textit{the parameters $\kappa$ and $\ell$ are completely independent of $\tau_{m_{*},n_{*}}$,  $u$ and  $v$}.\\ 
\indent Now, observe that $U, V, \tilde{U},\tilde{V}\neq 0$ (see \textit{Proposition 7} in \textit{Appendix Α}). Further, since $ U\tilde{V}-\tilde{U}V= \kappa-\ell \neq0$ (by \textit{Proposition 8} in \textit{Appendix Α}), it is easily seen that, for this option, the unique solution of the system $\left( \mathbb{S}_{\mathbb{Ζ}}\right)$ is exactly \\

\indent \hspace{3cm} $(X,Y,Z)=(x^{p-2},y^{q-2},z^{r-2})$. \\

But, on the other hand, considering the above-mentioned expression for this unique solution, we also get \\

\indent \hspace{3mm}$X=x^{p-2}$, $Y=\left(-z^{2}\right)D+ \left( x^{p} \tau_{m^{*},n^{*}} \right) E$ and $Z=(y^{2}) D + \left(x^{p}\right) E$ \\

\noindent (see \textit{Proposition 9} in the \textit{Appendix}), where we have used the notation
\begin{center}
$D:=\frac{\tau_{m_{*},n_{*}}z^{r-2} - y^{q-2}} {z^{2}- y^{2}\tau_{m_{*},n_{*}}}$ and $E:=\frac{1}{z^{2} - y^{2}\tau_{m_{*},n_{*}}}$.  
\end{center}  
In other words, adopting this formulation, we found
\begin{center}
$y^{q-2}=( -z^{2}) D+\left( x^{p} \tau_{m_{*},n_{*}} \right) E$ and $z^{r-2}=( - y^{2}) D + \left( x^{p}\right) E$. 
\end{center}
\noindent Since, from our assumption, $x^{p} =z^{r} - y^{q}$, it follows immediately that the system 
\[ 
\left\{\begin{array}{ll}
 \ \left(\tau_{m_{*},n_{*}} E z^{2} \right) Z +\left( -E \tau_{m_{*},n_{*}} y^{2} - 1 \right)  Y=  D z^{2}\\
 \left( 1-Ez^{2} \right) Z + \left(Ey^{2}\right) Y= - Dy^{2} 
\end{array}\right.
\]
\noindent would have the integer solution $ \left( Z,Y \right) =\left( z^{r-2}, y^{q-2} \right) \in \mathbb{Z}^{2}$. But, this is impossible since the determinant of this system is zero and both straight lines representing it are not parallel ( otherwise, the prime number $z^{2} - y^{2}\tau_{m_{*},n_{*}}$ would be written as the product of two integer factors: $z^{2} - y^{2}\tau_{m_{*},n_{*}}= y^{2}\left( -C- \tau_{m_{*},n_{*}} \right) $ for some $C \in \mathbb{Z} $, which is absurd; see \textit{Proposition 10} in \textit{Appendix Α}).  We therefore  conclude that there is no non-trivial solution to the Diophantine equation  $x^{p} + y^{q}=z^{r}$. $\square $ 

\begin{corol}{\normalfont(\textit{Fermat's Last Theorem})} 
If $ p \geq 3$, there is no non-trivial positive solution to the  Fermat equation $x^{p}+y^{p}=z^{p}$, provided that $gcd \left(  x, y, z \right)  =1$. $\square$
\end{corol}

\begin{rem}
{\normalfont The necessary condition $gcd(x,y,z)=1$ is essential, because otherwise the Diophantine equation $x^{p} + y^{q}=z^{r}$ may have infinitely many integer solutions. If, for instance, the exponent $r$ is any common multiple of $p$ and $q$ increased by one, say $r=\mu p+1=\eta q+1$, this equation has non-trivial integer solutions of the form $x=a(x^{p}+y^{q})^{\mu}$, $y=b(x^{p}+y^{q})^{\eta}$ and $z=x^{p}+y^{q}$, whenever $a$, $b$, $c\in\mathbb{Z}$, even if $p$ (either $q$) equals $2$ ([13]).} $\square $
\end{rem}

\begin{rem}
{\normalfont Possibly, the reader has been given the wrong impression that the proof of \textit{Theorem 1} is based only on arguments derived from Linear Algebra. Perhaps it has not yet become clear at all where the fact that the numbers $x$, $y$ and $z$ are three integers comes into play. So, a spontaneous question that may arise, at first glance, is where did we used the integrity of $x$, $y$ and $z$ in the proof of \textit{Theorem 1}. And where does our argument break down if $x$, $y$, $z$ is any real or complex solution? (Let us take, for example, in place of $\tau_{m_{*},n_{*}}$, any real or complex number $\tau$ for which the quantity $E=(z^{2}-y^{2} \tau)^{-1}$ can be well defined.)\\
\indent Unlike to the case of integers, any attempt to extend the same reasoning into a wider set $\mathbb{K}$, as $\mathbb{R}$  or $\mathbb{C}$, is doomed to fail, due to the fact that implementation of the prerequisite $\tau z^{r-2} - y^{q-2}\neq 0$  within $\mathbb{K}$ would lead to inability of defining the quantities  $U$, $V$, $\tilde{U}$ and $\tilde{V}$ as in the proof of \textit{Theorem 1}. To see this, let us assume that the three numbers $x$, $y$, $z \in \mathbb{K}\setminus \lbrace 0 \rbrace$ meet the equation $x^{p}+y^{q}=z^{r}$, for some $p$, $q$, $r \geq 3$. Suppose a number $\tau \in \mathbb{K}$ satisfies  $\tau z^{r-2} - y^{q-2}\neq 0$  and  $z^{2}y^{2}\tau\neq0$ and let  $(u,v)\in\mathbb{K}^{2}$ be any point of the line $(\epsilon):\left(x^{p-2}\right)u+ \left( \tau z^{r-2} -y^{q-2} \right) v = 1$. As in the proof of \textit{Theorem 1}, select again $U:=u+\kappa\left( \tau z^{r-2}-y^{q-2} \right)$, $V:=v-\kappa x^{p-2}$, $\tilde{U}:=u + \ell \left( \tau z^{r-2}-y^{q-2} \right)$ and $\tilde{V}:=v-\ell x^{p-2}$ with $\kappa$, $\ell \in \mathbb{K}$ ($\kappa\neq \ell$) and such that $U\tilde{V}- \tilde{U}V\neq 0$. It is necessary for our purposes to observe that \textit{the parameter $\kappa$ must be completely independent of $\tau$ and $u$(and also of $\ell$ and $v$)}. However, solving the corresponding system $\left( \mathbb{S}_{\mathbb{K}}\right)$ in $\mathbb{K}$, our assumption $\tau z^{r-2} - y^{q-2}\neq 0$ would impose dependence of $\kappa$ on $\tau$ and $u$. Indeed, since $U\tilde{V}- \tilde{U}V= \kappa -\ell \neq 0$, the (unique) solution of the system $\left( \mathbb{S}_{\mathbb{K}}\right)$ would be again $\left( X,Y,Z\right) =\left( x^{p-2},y^{q-2},z^{r-2} \right)$, and, dividing the first equation of $\left( \mathbb{S}_{\mathbb{K}}\right)$ with $V$, the second with $  \tilde{V}$ and the third one with  $z^{2}$, we would have
\[ 
\left\{\begin{array}{ll}
 \ \left( - \frac{U}{V} \right) X + Y + \frac{1}{V}= \tau Z\\
 \ \left( - \frac{\tilde{U}}{\tilde{V}} \right) X + Y + \frac{1}{\tilde{V}}= \tau Z\\
 \ \left(  \frac{x^{2}}{z^{2}} \right) X + \left(  \frac{ y^{2}}{z^{2}} \right) Y =  Z\\ 
\end{array}\right.
\Leftrightarrow
\left\{\begin{array}{ll}
 \ \left( - \frac{U}{V} \right) X  + \frac{1}{V}= \left( - \frac{\tilde{U}}{\tilde{V}} \right) X  + \frac{1}{\tilde{V}}\\
 \ \left( - \frac{U}{V} \right) X + Y + \frac{1}{V}= \left( \tau \frac{ x^{2}}{z^{2}} \right) X + \left( \tau \frac{ y^{2}}{z^{2}} \right) Y\\
 \ \left(  \frac{ x^{2}}{z^{2}} \right) X + \left(  \frac{ y^{2}}{z^{2}} \right) Y =  Z.\\ 
\end{array}\right.
\]
As $X=-\left( V-\tilde{V}\right)/ \left( U\tilde{V}-\tilde{U}V\right) =x^{p-2}$, the second equation would give 
\begin{center}
$ Y=\frac{z^{2}}{z^{2}-\tau  y^{2}} \left\lbrace  \left( \frac{U}{V}+\tau\frac{ x^{2}}{z^{2}}\right)  x^{p-2} - \frac{1}{V} \right\rbrace $.  
\end{center}
So, from $U:=u+\kappa \left( \tau z^{r-2}-y^{q-2} \right)$ and $V:=v-\kappa x^{p-2}$, we would get
\begin{center}
$ Y=\frac{z^{2}}{z^{2}-\tau  y^{2}} \left\lbrace  \left( \frac{u+\kappa\left( \tau z^{r-2}-y^{q-2} \right)}{v-\kappa x^{p-2}}+\tau\frac{ x^{2}}{z^{2}}\right)  x^{p-2} - \frac{1}{v-\kappa x^{p-2}} \right\rbrace $.  
\end{center}
Making simple algebraic operations, we would have\\

\indent $ Y=y^{q-2}=\frac{1}{\left( z^{2}-\tau  y^{2}\right) \left( 1-u x^{p-2} - \kappa x^{p-2} \left[ \tau z^{r-2} - y^{q-2} \right] \right)  }$ \\
\indent \hspace{2.6cm} $ (  \tau u x^{2p-2} +  \tau u x^{p-2} y^{q} - u x^{p-2} y^{q-2} z^{2} +  \kappa \tau^{2} x^{p-2} z^{2r-2}$\\
\indent \hspace{2.7cm}$-2  \kappa \tau x^{p-2} y^{q-2} z + \kappa x^{p-2} y^{2q-4} z^{2}-2  \tau x^{p}- \tau y^{q}$\\
\indent \hspace{2.9cm} $- \kappa \tau^{2} x^{2p-2} z^{r-2} +  \kappa \tau x^{2p-2} y^{q-2} +  y^{q-2} z^{2})$, \\
 
\noindent whenever $\kappa$, $u\in \mathbb{K}$ and $\tau \in \mathbb{K}$ (such that the denominator is $\neq 0$), and, from this, we would get the following representation of $\kappa $ as a function of $\tau$ and $u$
\begin{center}
$\kappa =\frac{x^{p}\left( u x^{p-2} -2 \right)}{\left( x^{2p-2} z^{r-2}+x^{p-2} y^{q}  z^{r-2} \right)\tau -  x^{2p-2} y^{q-2}-x^{p-2} y^{2q-2}-x^{p-2} y^{q-2} z^{r}} $,  
\end{center}
which would be impossible to happen, because the parameter $\kappa$ must have been taken to be independent of $\tau$ and $u$. 
$\square $
}
\end{rem}

In support of those mentioned above, let me point out that in $\mathbb{Z}$ the specific selection of the parameter $\tau_{m^{*},n^{*}}$ within the set of prime integers is essentially compatible for our purposes, since, besides other crucial properties, this particular choice meets the two necessary conditions
\begin{center}
 $D=( \tau_{m^{*},n^{*}} z^{r-2} - y^{q-2}) / (  z^{2} - \tau_{m^{*},n^{*}}  y^{2})  \neq 0$ 
\end{center} 
and $E=1 /(z^{2}- \tau_{m^{*},n^{*}}  y^{2} )< \infty$.\\
\indent In the context of integers, it is impossible (: in fact, it is "\textit{forbidden}") to exist a value for the parameter $\tau_{m^{*},n^{*}} \in \mathbb{Z}$ satisfying $D=0$, since any such option would imply $y^{q-2}=\tau_{m^{*},n^{*}} z^{r-2}$, and, thus, the integer $z$ would divide the integer $y$, in contrast to the straightforward consequence $gcd(y,z)=1$ of our basic hypothesis $gcd(x,y,z) = 1$ (see \textit{Proposition 1.i} in \textit{Appendix}).\\
\indent Interpreting the comparison, one can say that in fact \textit{the flexibility provided by the "free choice location" of the parameter $\tau$ everywhere within the set $\mathbb{K}\setminus \lbrace y^{q-2}/z^{r-2}\rbrace $ $\left( \mathbb{K}=\mathbb{R},\mathbb{C} \right)$ is drastically reduced if $\mathbb{K}=\mathbb{Z}$ and $\tau \in \mathbb{Z}\setminus \lbrace 0, \pm1\rbrace$, due to the role of $\tau=\tau_{m^{*},n^{*}}$ in the parametrization of the equation $\left(x^{p-2}\right)u+ \left( \tau_{m^{*},n^{*}} z^{r-2} -y^{q-2} \right) v = 1$} (: the case where one of the numbers $y$, $z$ is equal to $1$ has already been studied in [16]).\\
\indent  This drastic restriction disappears if $\mathbb{K}=\mathbb{R},\mathbb{C}$, since then, given any two numbers $y$, $z \in \mathbb{K} \setminus \lbrace 0,\pm1 \rbrace$  and any two values of the numerical parameters $q$, $r\in \mathbb{K} \setminus \lbrace 0 \rbrace$ satisfying $z^{r}\neq y^{q}$, it is always possible to find a $\tau\in \mathbb{K}$ such that 
\begin{center}
$y^{q-2}=\tau z^{r-2}$ ($ \Leftrightarrow D=0$)  
\end{center}

\noindent  with $z^{2}\neq \tau  y^{2}$ ($ \Leftrightarrow E\in \mathbb{K}$).  Indeed, observe that, due to the inequality $z^{r}\neq  y^{q}$, we can also write  $z^{2}\neq \left[  y^{q-2} / z^{r-2} \right] y^{2}$. Defining $\tau:=y^{q-2}/ z^{r-2}$, it is clear that $z^{2}\neq \tau y^{2}$ and $E\in \mathbb{K}$, in such a way that $ y^{q}=\tau z^{r-2} y^{2}$. Hence $z^{r}-  y^{q}=z^{r}-\tau z^{r-2} y^{2}=z^{r-2} \left[ z^{2}-\tau y^{2} \right] = z^{r-2}/Ε$, i.e., we conclude that $ z^{r}- y^{q}=z^{r-2}/Ε$. This, in particular, implies that, for each $p\in \mathbb{K}\setminus \lbrace 0 \rbrace$, the number $ x:=\left(1 / A \right)^{1/p} \left( z^{r-2}/Ε\right)^{1/p} \in \mathbb{K}\setminus \lbrace 0,\pm1 \rbrace$ , together with $y$, $z\in \mathbb{K}$, satisfy the given equation $x^{p}+y^{q}=z^{r}$.\\

\begin{rem} {\normalfont The above discussion can be summarized in \textit{a general methodological framework of finding a typical solution $\left( x,y,z\right)$ of the equation $x^{p}+y^{q}=z^{r} $ in $\mathbb{K}^{3}$} ($\mathbb{K}=\mathbb{R},\mathbb{C}$), in accordance with the following four successive algorithmic steps:
\begin{itemize}
  \item \textbf{Step 1}: Given $z, y\in \mathbb{K} \setminus \lbrace 0,\pm 1 \rbrace $ and $q$, $r\in \mathbb{K} \setminus \lbrace 0 \rbrace$ such that $z^{r}\neq y^{q}$, take \\
\indent \hspace{2cm}   $E:=\left[ z^{2}-\tau  y^{2} \right]^{-1}$, with $\tau:=y^{q-2}/z^{r-2}$ ($ \Leftrightarrow D=0$).\\
  \item \textbf{Step 2}: Since $y^{q-2}=\tau z^{r-2}$, the following typical relationship applies: \\ 
\indent \hspace{4cm}     $z^{r}- y^{q}=z^{r-2}/Ε$.
  \item \textbf{Step 3}: For each $p\in \mathbb{K} \setminus \lbrace 0 \rbrace $, define $ x:=\left( z^{r-2}/Ε\right)^{1/p}$. 
  \item \textbf{Step 4}: The numbers $x$, $y$ and $z$ satisfy  $x^{p}+y^{q}=z^{r} $ in $\mathbb{K}^{3} \setminus \lbrace 0 \rbrace$. $\square $
\end{itemize}
}
\end{rem}

\vspace{0.1mm}

\section{Applications to the Discrete Logarithm Problem}

Even if the confirmation of Beal's conjecture were to be seen as a negative result, there are some implications of it that could be mentioned. Let 's list a few examples on the discrete logarithm problem and its various versions.

\begin{corol} {\normalfont\textbf{(Non-existence of a solution to the discrete logarithm problem)}} 
Given any odd natural number $z$, there is no solution $r\geq 3$ of the discrete logarithm problem\\
\indent \hspace{5cm}     $z^{r}=t$,\\
if the natural number $t\in \mathbb{N}$ is written in the form\\
\indent \hspace{4,6cm} $t= x^{p}+ y^{q}$, \\
whenever $x$, $y$, $p$, $q\in \mathbb{N}$ such that $p\geq 3$ and $q\geq 3$ and $gcd (x, y, z) =1$. $\square $
\end{corol}

\begin{corol} {\normalfont\textbf{(Non-existence of a solution to the double discrete logarithm problem)}} 
For any odd natural number $M$, let us consider the equation\\
\indent \hspace{4,6cm}  $z= x^{p}+ y^{q}$ \\
with $x$, $y$, $p$, $q\in \mathbb{N}$ such that $p\geq 3$ and $q\geq 3$ and $gcd (x, y, z) =1$.\\
{\normalfont\textbf{i.}} There is no solution $r\geq 3$ of the double discrete logarithm problem\\
\indent \hspace{5cm}     $z= (M^{m})^{r}$,\\ 
whenever $m\in \mathbb{N}$ is taken to be such that  $gcd (x, y, M^{m}) =1$.\\
{\normalfont\textbf{ii.}} If, moreover, $m\geq 2$, then there is no integer solution $r\geq 2$ of the double discrete logarithm problem\\
\indent \hspace{5cm}     $z=M^{m^{r}}$,\\ 
such that $gcd (x, y, M) =1$. $\square $
\end{corol}

\begin{corol} {\normalfont\textbf{(Non-existence of a solution to the two-dimensional discrete logarithm problem)}} 
Let $m$ be any natural number. Let also $x$, $y$ and $z$ be any three natural numbers such that\\
\[gcd (x, y, M) =1 \hspace{1cm} and \hspace{1cm}  
\left\{\begin{array}{ll}
 \ x= even\\
 \ y= odd\\
 \ z= odd. 
\end{array}\right.
\]
Set\\
\indent \hspace{4.5cm}     $M_{x}:=m^{x}$,\\
\indent \hspace{4,5cm}     $M_{y}:=m^{y}$ and\\
\indent \hspace{4.5cm}     $M_{z}:=m^{z}$.\\

Then, for any integer $r\geq 3$, there is no solution pair $(p, q)$ of the two-dimensional discrete logarithm problem\\

\indent \hspace{4.5cm} $M_{x}^{p} M_{y}^{q}=M_{z}^{r} $\\
satisfying\\
\indent \hspace{5cm} $p$,$q\geq 3$. $\square $
\end{corol}

\vspace{0.2mm}

\begin{corol} {\normalfont\textbf{(Non-existence of a solution to the two-dimensional variant of the discrete logarithm problem)}} 
Given any two positive integers\\
\indent \hspace{4.5cm} $x=even$ and $y=odd$,\\
\noindent and a $t\in \mathbb{N}$, there is no solution pair $(p, q)$ of the two-dimensional variant of the discrete logarithm problem\\

\indent \hspace{4.5cm} $x^{p}+y^{q} = t $\\

\noindent whenever $t$ is a perfect power of degree greater than or equal to $3$, i.e. if\\

\indent \hspace{4.5cm} $t=z^{r}$\\

\noindent for $z \in \mathbb{N}$ and $r\in \mathbb{N}$ with $r\geq 3$. $\square $
\end{corol}

\vspace{0.3mm}

\section{On the Solution of Beal's Equation into Finite Fields}

We will now turn to the study of a ``local version" of the Beal equation, considering the (so-called) \textbf{\textit{Beal-Schur congruence equation}} $x^{p}+y^{q}\equiv z^{r} (mod  \mathcal{Ν})$. We will show that this  equation has non-trivial solution into the finite field $\mathbb{Z}_{\mathcal{Ν}}$ for all primes $\mathcal{Ν}$ that are sufficiently large. Specifically, adopting a proof structured by analogy with that given by Summer Lynne Kisner in [31], we will show the following result, which could be considered as an extension of Schur's Theorem, published in 1917 ([47]). It was introduced as a lemma in a paper meant to improve some results of Leonard Eugene Dickson and an English translation of the original proof appeared in [32].\\

\begin{theorem} 
Let $p$, $q$,$r$ be natural numbers that are greater than $1$ and such that \\
\indent \hspace{5cm} $q \mid r$ or $r \mid q$\\
and\\
\indent \hspace{5cm} $p \mid r$ or $r \mid q$\\
and\\
\indent \hspace{5cm} $q \mid p$ or $p \mid q$.\\
\noindent There is a natural number $\sigma = \sigma(p,q,r)$ such that for all primes $\mathcal{Ν} > \sigma $, the congruence \\
\indent \hspace{4.5cm} $x^{p}+y^{q}\equiv z^{r} (mod  \mathcal{Ν})$ \\
has a solution in the positive integers $x$, $y$, $z$ such that $\mathcal{Ν}$ does not divide $xyz$.
\end{theorem}
\indent The condition $\mathcal{Ν}$ does not divide $xyz $ is to avoid trivial solutions of the congruence, such as
$x \equiv y \equiv z \equiv 0$ or $x \equiv 0$, $y \equiv z (mod \mathcal{Ν})$.\\

\noindent \textit{Proof}.  The proof we will adopt is based on a Fourier approach in the additive group $\mathbb{Z}_{\mathcal{Ν}}$ and is structured by analogy with the one given by Summer Lynne Kisner ([15]. Let us consider the \textit{exponential sum}:\\

\indent \hspace{3.7cm} $\mathcal{S}_{k}^{(\ell)}= \sum_{x=0}^{\mathcal{Ν}-1} e^{2\pi i \frac{k x^{\ell} }{\mathcal{Ν}}}$\\

\indent \hspace{4.5cm} $= e^{2\pi i \frac{k }{\mathcal{Ν}}0^{\ell}} + e^{2\pi i \frac{k }{\mathcal{Ν}}1^{\ell}}+...+e^{2\pi i \frac{k }{\mathcal{Ν}}\left( \mathcal{Ν}-1\right) ^{\ell}}  $.\\

Let also \\
\indent \hspace{4.7cm} $\mathcal{M}$ \\

\noindent denote the number of ordered triples $(x, y, z)$ in $\mathbb{Z}_{\mathcal{Ν}}^{3}$ such that $x^{p}+y^{q}=z^{r}$. Note, that this happens precisely when \\

\indent \hspace{4cm} $x^{p}+y^{q}- z^{r} \equiv 0 (mod  \mathcal{Ν})$. \\

\noindent We shall now show that if $\mathcal{Ν}$ is sufficiently large, there exists a positive constant $s>0$ such that $\mathcal{M}>s$; in other words, if $\mathcal{Ν}>0$, there must be, at least, a non-trivial solution to the equation $x^{p}+y^{q}=z^{r}$ in $\mathbb{Z}_{\mathcal{Ν}}$.\\

Since (by \textit{Lemma 3} in Appendix B), 
\begin{center}
$  \sum_{k\in \mathbb{Z}_{\mathcal{Ν}}} e^{2\pi i  \frac{k}{\mathcal{Ν}} \left( x^{p}+y^{q}-z^{r} \right) }= \sum_{k=0}^{\mathcal{Ν}-1} e^{2\pi i \frac{k}{\mathcal{Ν}}\left( x^{p}+y^{q}-z^{r} \right)} =  
\left \{
\begin{array}{ll}
 \ \mathcal{Ν}, \hspace{0.1cm} if \hspace{0.1cm}  x^{p}+y^{q} \equiv z^{r} \left( mod \mathcal{Ν} \right) \\
 \ 0, \hspace{0.1cm} otherwise 
\end{array} 
\right. $  
\end{center} 
\noindent we see that if the triplet $\left( x, y, z \right)$ is a solution to the equation $ x^{p}+y^{q} -z^{r} \equiv 0 \left( mod \mathcal{Ν} \right) $, then
\begin{center}
$ \frac{1}{\mathcal{Ν}} \sum_{k=0}^{\mathcal{Ν}-1} e^{2\pi i \frac{k}{\mathcal{Ν}}\left( x^{p}+y^{q}-z^{r} \right)} =  1 $  
\end{center} 
\noindent (and when it is not, then 
\begin{center}
$ \frac{1}{\mathcal{Ν}} \sum_{k=0}^{\mathcal{Ν}-1} e^{2\pi i \frac{k}{\mathcal{Ν}}\left( x^{p}+y^{q}-z^{r} \right)} =  0 $).  
\end{center} 

Based on this observation, we now have a direct way to find the number $ \mathcal{M}$ of solutions to the equation  $x^{p}+y^{q} \equiv z^{r} \left( mod \mathcal{Ν} \right)$ by setting:
\begin{center}
$ \mathcal{M} = \sum_{x, y, z \in \mathbb{Z}_{\mathcal{Ν}}} \frac{1}{\mathcal{Ν}} \sum_{k=0}^{\mathcal{Ν}-1} e^{2\pi i \frac{k}{\mathcal{Ν}}\left( x^{p}+y^{q}-z^{r} \right)} $.  
\end{center}
 
\noindent Hence, it suffices to analyse this summation expression that gives the number $\mathcal{M}$ to show that for any sufficiently large prime number $\mathcal{Ν}$, the equation under consideration has a non-trivial solution $\left( x, y, z \right)$. To do so, we may proceed as follows:\\

\indent \hspace{1.5cm} $\mathcal{M}= \frac{1}{\mathcal{Ν}} \sum_{k=0}^{\mathcal{Ν}-1} \sum_{x \in \mathbb{Z}_{\mathcal{Ν}}}  e^{ \frac{2\pi i k x^{p}}{\mathcal{Ν}}}
\sum_{y \in \mathbb{Z}_{\mathcal{Ν}}}  e^{ \frac{2\pi i k y^{q}}{\mathcal{Ν}}}
\sum_{z \in \mathbb{Z}_{\mathcal{Ν}}}  e^{ \frac{2\pi i k \left( -z^{r} \right) }{\mathcal{Ν}}}$\\

\indent \hspace{2.02cm} $=\frac{1}{\mathcal{Ν}} \sum_{k=0}^{\mathcal{Ν}-1} \mathcal{S}_{k}^{(p)} \mathcal{S}_{k}^{(q)} 
\conjug{\mathcal{S}_{k}^{(r)}}$\\

\indent \hspace{2.02cm} $=\frac{1}{\mathcal{Ν}} \left( \mathcal{S}_{0}^{(p)} \mathcal{S}_{0}^{(q)} \conjug{\mathcal{S}_{0}^{(r)}} + \sum_{k=1}^{\mathcal{Ν}-1} \mathcal{S}_{k}^{(p)} \mathcal{S}_{k}^{(q)} \conjug{\mathcal{S}_{k}^{(r)}} \right)  $\\

\indent \hspace{2.02cm} $=\frac{1}{\mathcal{Ν}} \left( \mathcal{N}^{3} + \sum_{k=1}^{\mathcal{Ν}-1} \mathcal{S}_{k}^{(p)} \mathcal{S}_{k}^{(q)} \conjug{\mathcal{S}_{k}^{(r)}} \right)  $\\

\indent \hspace{2.02cm} $=\mathcal{N}^{2} + \frac{1}{\mathcal{Ν}} \sum_{k=1}^{\mathcal{Ν}-1} \mathcal{S}_{k}^{(p)} \mathcal{S}_{k}^{(q)} \conjug{\mathcal{S}_{k}^{(r)}} $.\\

\noindent From the fact that the difference\\
\indent \hspace{5.3cm} $ \mathcal{M} - \mathcal{N}^{2} $\\
\noindent is an integer and, in particular, a real number, it follows that the sum\\

\indent \hspace{4.3cm} $\frac{1}{\mathcal{Ν}} \sum_{k=1}^{\mathcal{Ν}-1} \mathcal{S}_{k}^{(p)} \mathcal{S}_{k}^{(q)} \conjug{\mathcal{S}_{k}^{(r)}} $\\

\noindent is also a real number. This implies that:\\

\noindent \hspace{1cm} $\frac{1}{\mathcal{Ν}} \sum_{k=1}^{\mathcal{Ν}-1} \mathcal{S}_{k}^{(p)} \mathcal{S}_{k}^{(q)} \conjug{\mathcal{S}_{k}^{(r)}} \geq -\vert \frac{1}{\mathcal{Ν}} \sum_{k=1}^{\mathcal{Ν}-1} \mathcal{S}_{k}^{(p)} \mathcal{S}_{k}^{(q)} \conjug{\mathcal{S}_{k}^{(r)}} \vert $\\

\indent \hspace{4cm} $\geq - \frac{1}{\mathcal{Ν}} \sum_{k=1}^{\mathcal{Ν}-1} \vert \mathcal{S}_{k}^{(p)} \mathcal{S}_{k}^{(q)} \conjug{\mathcal{S}_{k}^{(r)}} \vert $\\

\indent \hspace{4cm} $\geq - \frac{1}{\mathcal{Ν}} \sum_{k=1}^{\mathcal{Ν}-1} \vert e^{2\pi i \frac{k x^{p}}{\mathcal{Ν}}} \vert \vert e^{2\pi i \frac{k y^{q}}{\mathcal{Ν}}} \vert \vert e^{-2\pi i \frac{k z^{r}}{\mathcal{Ν}}} \vert$\\

\indent \hspace{4cm} $\geq - \frac{1}{\mathcal{Ν}} \sum_{k=1}^{\mathcal{Ν}-1} \left( \sqrt{2 \mathcal{Ν}} p \right) \left( \sqrt{2 \mathcal{Ν}} q \right) \left( \sqrt{2 \mathcal{Ν}} r \right)$\\

\indent \hspace{6cm} (by \textit{Proposition 11} in Appendix B)\\

\indent \hspace{4cm} $=-\frac{\mathcal{Ν}-1}{\mathcal{Ν}} \left(  2 \mathcal{Ν} \right)^{\frac{3}{2}} p q r $\\

\indent \hspace{4cm} $\geq -\left(  2 \mathcal{Ν} \right)^{\frac{3}{2}} p q r $.\\

\noindent Thus, we have proved that
\begin{equation}
\mathcal{M} \geq \mathcal{Ν}^{2}-\left(  2 \mathcal{Ν} \right)^{\frac{3}{2}} p q r.
\end{equation}

\indent Now, suppose that the prime number $\mathcal{Ν}$ is chosen large enough so that
\begin{equation}
\mathcal{N} \geq \sigma = \sigma(p,q,r):= 32 p^{2} q^{2} r^{2}.
\end{equation}
Having made this choice, we observe that the prime $\mathcal{Ν}$ must satisfy the following inequality which is equivalent to that of $(2)$:
\begin{center}
$ \frac{1}{4} \mathcal{Ν}^{4} \geq 8 p^{2} q^{2} r^{2} \mathcal{Ν}^{3}.$  
\end{center} 
This obviously amounts to the inequality $ \frac{1}{2} \mathcal{Ν}^{2} \geq 2^{\frac{3}{2}} p q r \mathcal{Ν}^{\frac{3}{2}}$ or, even, to the inequality
\begin{equation}
\mathcal{Ν}^{2}-\left(  2 \mathcal{Ν} \right)^{\frac{3}{2}} p q r \geq \frac{1}{2} \mathcal{Ν}^{2}.
\end{equation}
So, combining (3) with (1), we get
\begin{equation}
\mathcal{M} \geq \frac{1}{2} \mathcal{Ν}^{2}.
\end{equation}

But, on the other hand, it is easy to see that
\begin{equation}
\frac{1}{2}\mathcal{Ν}^{2}>1+(p+q+r)\mathcal{Ν}.
\end{equation}
\noindent Indeed, by (2), the prime  $ \mathcal{Ν}$ must be chosen so that $\mathcal{N} \geq 32 p^{2} q^{2} r^{2}$. Obviously, this inequality holds if and only if $\frac{\sqrt{\mathcal{N}}}{\sqrt{32}}>pqr$ (: the inequality is strict, since $\mathcal{Ν}$ is a prime number, so it cannot be equal to $32 p^{2} q^{2} r^{2}$). Since $pqr>p+q+r$, we infer $ \frac{\sqrt{\mathcal{N}}}{\sqrt{32}} > p+q+r$ which can also be written in the form
\begin{equation}
7 \frac{\sqrt{\mathcal{N}}}{\sqrt{32}} \mathcal{Ν} > 7 (p+q+r) \mathcal{Ν}.  
\end{equation}
Since, by (2), 
\begin{equation}
\frac{1}{2}\mathcal{Ν}^{2} >7 \frac{\sqrt{\mathcal{N}}}{\sqrt{32}} \mathcal{Ν},  
\end{equation}
and 
\begin{equation}
7 (p+q+r) \mathcal{Ν}>1+ (p+q+r) \mathcal{Ν},  
\end{equation}
combination of (6), (7) and (8) proves inequality (5). \\

\indent Therefore, combining (4) and (5), we conclude that
\begin{equation}
\mathcal{M} > 1+ (p+q+r) \mathcal{Ν}.  
\end{equation}
\indent Now, to complete the proof, it suffices to count the number of trivial solutions and show that it is less than $1+ (p+q+r) \mathcal{Ν}$. To this end, observe that trivial solutions to the equation $x^{p}+y^{q}\equiv z^{r} (mod  \mathcal{Ν})$ are those where $\mathcal{Ν}$ divides $xyz$. Note that when $x = y = z = 0$, we have a trivial solution. So we have at least a trivial solution and we need to count those where exactly one of $x$, $y$ or $z$ is zero (because if any two are zero the last will be forced to be zero.) This means we have the following three cases. \\

\indent If $r \mid q $ and $x = 0$, the number of solutions is exactly the number of pairs with $y^{q}\equiv z^{r} \left( mod \mathcal{N}\right)$ with $y$, $z \neq 0$. But, whenever $z\in \mathbb{Z}_{\mathcal{Ν}}^{*}$ is fixed, this number is the same as the number of pairs $\left( tz, z \right)$ where $t^{r}=1 $ that is\\
\begin{equation}
t=e^{\frac{2k \pi i}{r}} (k=0,1,...,r-1).  
\end{equation}
Since there are at most $\mathcal{Ν}-1$ choices for $z$ and at most $r$ solutions for the equation (10), we are led to the conclusion that there are at most\\
\indent \hspace{5.1cm}$r \left( \mathcal{Ν}-1 \right)$\\
solutions to the equation $y^{q}\equiv z^{r} (mod  \mathcal{Ν})$ whenever $z\in \mathbb{Z}_{\mathcal{Ν}}^{*}$ is fixed. Similarly, if $x=0$,  $y\in \mathbb{Z}_{\mathcal{Ν}}^{*}$ is fixed and $q \mid r $, we see that there are at most\\
\indent \hspace{5.1cm}$q \left( \mathcal{Ν}-1 \right)$\\
solutions to the equation $y^{q}\equiv z^{r} (mod  \mathcal{Ν})$. In summary, under the stated assumptions, the number of solutions to the equation $y^{q}\equiv z^{r} (mod  \mathcal{Ν})$ is less than\\
\indent \hspace{4.7cm}$ min \left\lbrace q, r\right\rbrace  \left( \mathcal{Ν}-1 \right)$.\\
\indent By symmetry, the analogous upper bound applies to the number of solutions when $y = 0$:\\
\indent \hspace{4.7cm}$min \left\lbrace p, r\right\rbrace  \left( \mathcal{Ν}-1 \right)$\\
and by a similar argument, it also applies to the upper bound for the number of solutions when $z = 0$:\\
\indent \hspace{4.7cm}$min \left\lbrace p, q\right\rbrace  \left( \mathcal{Ν}-1 \right)$\\
(here, we may consider the equations\\
\indent \hspace{1.8cm}
$t=e^{-\frac{2k \pi i}{q}} (k=0,1,...,q-1)$ and $t=e^{-\frac{2k \pi i}{p}} (k=0,1,...,p-1),$\\
instead of the equations\\
\indent \hspace{1.8cm} 
$t=e^{\frac{2k \pi i}{q}} (k=0,1,...,q-1)$ and $t=e^{\frac{2k \pi i}{p}}  (k=0,1,...,p-1)$.)\\

Thus, we have showed that \textit{the number of trivial solutions is at most} 
\begin{center}
$1+ \left( min \left\lbrace q, r\right\rbrace +min \left\lbrace p, r\right\rbrace +min \left\lbrace p, q\right\rbrace \right)  \left( \mathcal{Ν}-1 \right)$. 
\end{center}
But, it is clear that \textit{this upper bound for the number of trivial solutions  of the equation $x^{p}+y^{q} \equiv z^{r} \left( mod \mathcal{Ν}\right) $  is less than the number $1+ (p+q+r) \mathcal{Ν}$ which is a lower bound for the number $\mathcal{M}$ of all solutions (trivial and non trivial) of the given equation}:
\begin{center}
$\mathcal{M}>1+ (p+q+r) \mathcal{Ν}>1+ \left( min \left\lbrace q, r\right\rbrace +min \left\lbrace p, r\right\rbrace +min \left\lbrace p, q\right\rbrace \right)  \left( \mathcal{Ν}-1 \right)$. 
\end{center}

This means we have proved that there is at least one non-trivial solution to the equation $x^{p}+y^{q} \equiv z^{r} \left( mod \mathcal{Ν}\right) $, in the sense  that from some point on, namely when
\begin{center}
$\mathcal{N} \geq \sigma = \sigma(p,q,r):= 32 p^{2} q^{2} r^{2}$, 
\end{center}
there must be a non-trivial solution to the equation $x^{p}+y^{q}\equiv z^{r}$  in $\mathbb{Z}_{\mathcal{Ν}}$.  $ \square $\\

\vspace{1mm}

\section{BS Cryptosystems}

As an implementation of the above, we will now give three algorithms that introduce post-quantum encryption schemes based on solving multi-variate Diophantine congruence equations, and, in particular,  the Beal-Schur congruence equation. We call these algorithms \textbf{\textit{BS cryptosystems}}. Cryptanalysis of the corresponding mathematical models demonstrates the potential of applying this method, since the use of such a congruence equation allows the construction of systems which have a large variety of equally probable keys (actually, there is an infinite number of key's choices). And only one key is correct...\\
\indent To ensure uniformity in the terminology we will adopt, we introduce the following definitions.

\begin{defi}  
A triplet $\left( p, q, r \right)$ of natural numbers that are greater than $1$ is called intra-divisible, if the following relationships hold: \\
\indent \hspace{5cm} $q \mid r$ or $r \mid q$\\
and\\
\indent \hspace{5cm} $p \mid r$ or $r \mid q$\\
and\\
\indent \hspace{5cm} $q \mid p$ or $p \mid q$. $\square $
\end{defi} 

\begin{defi}  
Given an intra-divisible triplet $\left( p, q, r \right)$, a prime number $\mathcal{Ν}$ is an indiscernible large prime with respect to the triplet $\left( p, q, r \right)$ if 
\begin{center}  
$\mathcal{Ν}> 32p^{2}q^{2}r^{2}$. $\square $
\end{center} 
\end{defi} 

\begin{defi}  
Let $\left( p, q, r \right)$ be an intra-divisible triplet and $\mathcal{Ν}$ is an indiscernible large prime with respect to the triplet $\left( p, q, r \right)$. A triplet $\left( x, y, z \right)$ of natural numbers such that
\begin{center}  
$x^{p} + y^{q} \equiv z^{r} \left( mod \mathcal{Ν} \right) $
\end{center} 
is an $\mathcal{Ν}-$congruent Beal-Schur triplet with respect to the triplet $\left( p, q, r \right)$. $\square $
\end{defi}

\subsection{First Scheme of BS Cryptosystem}

\indent\hspace{1.6em}\texttt{Encryption I}
\begin{description}
\item[$\bullet$]\indent\hspace{0.3em} Let $r \in \mathbb{N}\setminus\left\lbrace 1 \right\rbrace $  (: the first component of a public key).\\
\item[$\bullet$]\indent \hspace{0.3em} Choose randomly a private key $\left( p, q \right) \in \left( \mathbb{N}\setminus\left\lbrace 1 \right\rbrace \right)^{2}$ so that the triplet $\left( p, q, r \right)$ is intra-divisible.\\
\item[$\bullet$]\indent \hspace{0.3em} Choose randomly an indiscernible large prime $\mathcal{N} \in \mathbb{N}$ (: the second component of the public key) with respect to the triplet $\left( p, q, r \right)$.\\
\item[$\bullet$]\indent \hspace{0,3em} Map the message $\mathfrak{M}$ to an element $z$ of $\mathbb{Z}_{\mathcal{N}}$ using a reversible mapping.\\
\item[$\bullet$]\indent \hspace{0.3em} Compute $x$, $y\in\mathbb{Z}_{\mathcal{N}}$ such that the triplet $\left( x, y, z \right)$ is an $\mathcal{Ν}-$congruent Beal-Schur triplet with respect to the triplet $\left( p, q, r \right)$, i.e.\\

\hspace{12em}  $x^{p} + y^{q} \equiv z^{r} \left( mod \mathcal{N} \right)$.\\

\item[$\bullet$]\indent \hspace{0.3em} The first party (Alice) sends the cipher text $\left( x, y \right)$ to a second party (Bob).

\end{description}

\indent\hspace{1.6em}\texttt{Decryption I}

\begin{description}
\item[$\bullet$]\hspace{0,3em} Given the public key $\left(r, \mathcal{N} \right)\in \left( \mathbb{N}\setminus \left\lbrace 1 \right\rbrace \right)^{2}$,\\

\indent \hspace{1em} if Bob knows the cipher text $\left( x, y \right)$ and the private key $\left( p, q \right) $,\\

\indent \hspace{1em} he can find the plain text $z$ by solving the equation: \\

\hspace{12em} $z^{r} \equiv x^{p} + y^{q} \left( mod \mathcal{N} \right)$.\\

\item[$\bullet$]\hspace{0,3em} Bob maps $z$  back to the plain text message  $\mathfrak{M}$.

\end{description}

\vspace{2mm}
 
\subsection{Second Scheme of BS Cryptosystem}

\indent\hspace{1.6em}\texttt{Encryption II}
\begin{description}
\item[$\bullet$]\indent\hspace{0.3em} Choose randomly an intra-divisible triplet $\left( p, q, r \right) \in \left( \mathbb{N}\setminus\left\lbrace 1 \right\rbrace \right)^{3} $  which is the public key.\\

\item[$\bullet$]\indent \hspace{0.3em} Choose randomly a private key to be an indiscernible large prime $\mathcal{N} \in \mathbb{N}$ with respect to the triplet $\left( p, q, r \right)$. \\

\item[$\bullet$]\indent \hspace{0,3em} Map the message $\mathfrak{M}$ to an element $z$ of $\mathbb{Z}_{\mathcal{N}}$ using a reversible mapping.\\

\item[$\bullet$]\indent \hspace{0.3em} Compute $x$, $y\in\mathbb{Z}_{\mathcal{N}}$ such that the triplet $\left( x, y, z \right)$ is an $\mathcal{Ν}-$congruent Beal-Schur triplet with respect to the triplet $\left( p, q, r \right)$, i.e.\\

\hspace{12em}  $x^{p} + y^{q} \equiv z^{r} \left( mod \mathcal{N} \right)$.\\

\item[$\bullet$]\indent \hspace{0.3em} The first party (Alice) sends the cipher text $\left( x, y \right)$ to a second party (Bob).

\end{description}

\indent\hspace{1.6em}\texttt{Decryption II}

\begin{description}
\item[$\bullet$]\hspace{0,3em} Given the public key $\left(p, q, r \right)\in \left( \mathbb{N}\setminus \left\lbrace 1 \right\rbrace \right)^{3}$,\\

\indent \hspace{1em} if Bob knows the cipher text $\left( x, y \right)$ and the private key $\mathcal{N} $,\\

\indent \hspace{1em} he can find the plain text $z$ by solving the equation: \\

\hspace{12em} $z^{r} \equiv x^{p} + y^{q} \left( mod \mathcal{N} \right)$.\\

\item[$\bullet$]\hspace{0,3em} Bob maps $z$  back to the plain text message  $\mathfrak{M}$.

\end{description}

\vspace{2mm}
 
\subsection{Third Scheme of BS Cryptosystem}

\indent\hspace{1.6em}\texttt{Encryption III}
\begin{description}
\item[$\bullet$]\indent \hspace{0,3em} Let $\mathfrak{M}=\mathfrak{M}_{1}\mathfrak{M}_{2}...\mathfrak{M}_{n}$ be a private key partition of the message $\mathfrak{M}$ to be transmitted.\\

\item[$\bullet$]\indent \hspace{0,3em} Choose randomly $n$ intra-divisible triplets $\left( p_{i}, q_{i}, r_{i} \right) \in \left( \mathbb{N}\setminus\left\lbrace 1 \right\rbrace \right)^{3} $ ($i=1, 2,..., n$) which are considered to be parts of the public key.\\

\item[$\bullet$]\indent \hspace{0.3em} Choose randomly $n$ private keys to be $n$ indiscernible large prime numbers $\mathcal{N}_{i} \in \mathbb{N}$ with respect to the triplets $\left( p_{i}, q_{i}, r_{i} \right)$. \\

\item[$\bullet$]\indent \hspace{0,3em} Map each message $\mathfrak{M}_{i}$ to an element $z_{i}$ of $\mathbb{Z}_{\mathcal{N}_{i}}$ using a reversible mapping. \\

\item[$\bullet$]\indent \hspace{0,3em} Let \\

\indent \hspace{10em} $i_{1}$, $i_{2}$,..., $i_{\lambda}$  (public or private key)\\

the indices for which \texttt{ENCRYPTION I} will be applied and let\\

\indent \hspace{10em} $i_{\lambda+1}$, $i_{\lambda+2}$,..., $i_{n}$  (private or public key, respectively)\\

the indices for which \texttt{ENCRYPTION II} will be applied.\\

\item[$\bullet$]\indent\hspace{0.3em} Apply \texttt{ENCRYPTION I} for the messages\\

\indent \hspace{10em}$\mathfrak{M}_{i_{1}}\mathfrak{M}_{i_{2}}...\mathfrak{M}_{i_{\lambda}}$\\

to compute $\left( x_{i_{1}}, y_{i_{1}} \right)$, ..., $\left( x_{i_{\lambda}}, y_{i_{\lambda}} \right)$ such that the triplets $\left( x_{i_{j}}, y_{i_{j}}, z_{i_{j}} \right)$ are $\mathcal{Ν}_{j}-$congruent Beal-Schur triplets with respect to the respective triplets $\left( p_{i_{j}}, q_{i_{j}}, r_{i_{j}} \right)$, i.e.\\

\hspace{12em}  $x_{i_{j}}^{p_{i_{j}}} + y_{i_{j}}^{q_{i_{j}}} \equiv z_{i_{j}}^{r_{i_{j}}} \left( mod \mathcal{N}_{i_{j}} \right)$\\

whenever $j=1, 2,..., \lambda $.\\

\item[$\bullet$]\indent\hspace{0.3em} Apply \texttt{ENCRYPTION II} for the messages\\

\indent \hspace{10em}$\mathfrak{M}_{i_{\lambda+1}}\mathfrak{M}_{i_{\lambda+2}}...\mathfrak{M}_{i_{n}}$\\

to compute $\left( x_{i_{\lambda+1}}, y_{i_{\lambda+1}} \right)$, ..., $\left( x_{i_{n}}, y_{i_{n}} \right)$ such that the triplets $\left( x_{i_{j'}}, y_{i_{j'}}, z_{i_{j'}} \right)$ are  $\mathcal{Ν}_{j'}-$congruent Beal-Schur triplets with respect to the respective triplets $\left( p_{i_{j'}}, q_{i_{j'}}, r_{i_{j'}} \right)$, i.e.\\

\hspace{9em}  $x_{i_{j'}}^{p_{i_{j'}}} + y_{i_{j'}}^{q_{i_{j'}}} \equiv z_{i_{j'}}^{r_{i_{j'}}} \left( mod \mathcal{N}_{i_{j'}} \right)$\\

whenever $j'= \lambda+1,...,n $.\\

\indent \hspace{11em}$\mathfrak{M}_{i_{\lambda+1}}\mathfrak{M}_{i_{\lambda+2}}...\mathfrak{M}_{n}$\\

\item[$\bullet$]\indent \hspace{0.3em} The first party (Alice) sends the cipher text \\

\indent \hspace{6em}$\left( x_{i_{1}}, y_{i_{1}} \right)$, ..., $\left( x_{i_{\lambda}}, y_{i_{\lambda}} \right)$, $\left( x_{i_{\lambda+1}}, y_{i_{\lambda+1}} \right)$, ..., $\left( x_{i_{n}}, y_{i_{n}} \right)$ \\

to a second party (Bob).

\end{description}

\vspace{3mm}

\indent\hspace{1.6em}\texttt{Decryption III}

\begin{description}
\item[$\bullet$]\hspace{0,3em} Given the public key $i_{1}$, $i_{2}$,..., $i_{\lambda}$ (or $i_{\lambda+1}$, $i_{\lambda+2}$,..., $i_{n}$),\\

\indent \hspace{1em} if Bob knows the cipher text\\
 
\hspace{8em} $\left( x_{i_{1}}, y_{i_{1}} \right)$, ..., $\left( x_{i_{\lambda}}, y_{i_{\lambda}} \right)$, $\left( x_{i_{\lambda+1}}, y_{i_{\lambda+1}} \right)$, ..., $\left( x_{i_{n}}, y_{i_{n}} \right)$ 

\indent \hspace{1em} and the private key\\

\hspace{10em}$\left(  i_{\lambda+1}, i_{\lambda+2},..., i_{n}, \mathcal{Ν}_{1}, \mathcal{Ν}_{2}, ...,\mathcal{Ν}_{n}\right) $ \\

\indent \hspace{1em} (or the private key\\

\hspace{11em}$\left(  i_{1}, i_{2},..., i_{\lambda}, \mathcal{Ν}_{1}, \mathcal{Ν}_{2}, ...,\mathcal{Ν}_{n}\right) $), \\

\indent \hspace{1em} he can find the plain texts:\\

\hspace{11em}  $z_{i}^{r_{i}} \equiv x_{i}^{p_{i}} + y_{i}^{q_{i}}  \left( mod \mathcal{N}_{i} \right)$\\

\indent \hspace{1em} whenever $i=1, 2,..., n $.\\

\item[$\bullet$]\hspace{0,3em} Bob maps the resulting cipher text \\

\hspace{14em} $z_{1}z_{2}...z_{n}$ \\

back to the plain text message \\

\hspace{12em}  $\mathfrak{M}_{1}\mathfrak{M}_{2}...\mathfrak{M}_{n} = \mathfrak{M}$.

\end{description}

\vspace{2mm}

\section{BS Key Generation Algorithmic Schemes}

Besides BS encryption algorithms, the application of Theorem 2 can yield simple and secure secret \textit{\textbf{BS}}(: for Beal-Schur) \textit{\textbf{key generation algorithms}}. Below, we will give two such post-quantum algorithmic schemes.

\subsection{First BS Key Generation Algorithmic Scheme}

The first party, Alice, generates a key as follows:

\begin{description}
\item[$\bullet$]\hspace{0,3em} Choose randomly an intra-divisible triplet $\left(p, q, r \right) \in \mathbb{Z}^{3}$ (: the first component of the private key).\\

\item[$\bullet$]\hspace{0,3em} Choose randomly an indiscernible large prime $\mathcal{N} \in \mathbb{N}$ (: the first component of the public key) with respect to the triplet $\left( p, q, r \right)$.\\

\item[$\bullet$]\hspace{0,3em} Given a natural number $z\in \mathbb{Z}_{\mathcal{N}}$ (: the second component of the public key), compute $\left( x, y \right) \in \mathbb{Z}_{\mathcal{N}}^{2}$ (: the second component of the private key), in such a way that the triplet  $\left( x, y, z \right) \in \mathbb{Z}_{\mathcal{N}}^{3}$ is an $\mathcal{Ν}-$congruent Beal-Schur triplet with respect to the triplet $\left( p, q, r \right)$, i.e. $ x^{p} + y^{q} \equiv z^{r}  \left( mod \mathcal{N} \right)$.\\

\item[$\bullet$]\hspace{0,3em} The public key consists of the values $\left( \mathcal{N}, z \right) $. Alice publishes this public key and retains \\
\indent \hspace{10em} $\left(p, q, r, x, y \right) \in \mathbb{Z}^{3}\times\mathbb{Z}_{\mathcal{N}}^{2}$ \\

as her private key, which must be kept secret.

\end{description}

\subsection{Second BS Key Generation  Algorithmic Scheme}

The first party, Alice, generates a key as follows:

\begin{description}
\item[$\bullet$]\hspace{0,3em} Choose randomly an intra-divisible triplet $\left(p, q, r \right) \in \mathbb{Z}^{3}$ (: the first component of the public key).\\

\item[$\bullet$]\hspace{0,3em} Choose randomly an indiscernible large prime $\mathcal{N} \in \mathbb{N}$ (: the first component of the private key) with respect to the triplet $\left( p, q, r \right)$.\\

\item[$\bullet$]\hspace{0,3em} Given a natural number $z\in \mathbb{Z}_{\mathcal{N}}$ (: the second component of the public key), compute $\left( x, y \right) \in \mathbb{Z}_{\mathcal{N}}^{2}$ (: the second component of the private key), in such a way that the triplet  $\left( x, y, z \right) \in \mathbb{Z}_{\mathcal{N}}^{3}$ is an $\mathcal{Ν}-$congruent Beal-Schur triplet with respect to the triplet $\left( p, q, r \right)$, i.e. $ x^{p} + y^{q} \equiv z^{r}  \left( mod \mathcal{N} \right)$.\\

\item[$\bullet$]\hspace{0,3em} The public key consists of the values $\left(p, q, r, x, y \right) \in \mathbb{Z}^{5}$. Alice publishes this key and retains \\
\indent \hspace{10em} $\left(\mathcal{Ν}, x, y \right) \in \mathbb{N}\times\mathbb{Z}_{\mathcal{Ν}}^{2}$  \\

as her private key, which must be kept secret.

\end{description}

\begin{rem}{\normalfont Of course, one could consider the congruence equation of another type of three-variable polynomial equation, quite different from a Beal-Schur Diophantine Equation, such as, for example, the Diophantine Equation}  $ y^{2} + z^{2} + 5 = x^{3} + xyz${\normalfont , which has no integer solution in $\mathbb{Z}$, as proved in the paper} ``Fruit Diophantine Equation" {\normalfont by Dipramit Majumdar and B. Sury (see https://arxiv.org/abs/2108.02640v2). Doing so might significantly increase the degree of encryption security.} $ \square $
\end{rem}

\begin{large}
\indent \hspace{4.5cm}\textbf{Appendix Α}
\end{large}%

\vspace{5mm}
 
\indent We will now give a series of ten auxiliary procedural effects having purely computational documentation that has been preferred not to be included in the main body of the text for reasons of structural aesthetics, but also to facilitate direct supervision of key arguments underlying the substantive steps of the proof in \textit{Theorem 1} as well as of the relevant material contained in subsequent \textit{Remarks}.\\
\indent Assume that there are three integer exponents $p$, $q$, $r\geq 3$ and three coprime positive integers $x=even$, $y=odd$ and $z=odd$ satisfying a generalized Fermat equation of the form  $x^{p}+ y^{q}=z^{r}$.

\begin{proposition}  The following three relationships apply:\\
\textbf{i}.  \hspace{0.2cm} $gcd \left( y, z \right) =1$.\\
\textbf{ii}. \hspace{0.1cm} $gcd \left(  y^{k}, z^{l} \right) =1$, where $k$ and $l$ are any two positive integers such that $k\leq q$ and $l\leq r $.\\
\textbf{iii}.  $gcd \left( - y^{q-2}, \left[ z^{2} - \tau_{0} y^{2} \right] \right) =1$.
\end{proposition}
\textit{Proof} \textbf{i}. Suppose that, on the contrary, we have $gcd \left( y, z \right) = \varepsilon >1$. Then $y=\varepsilon \Phi $ and $z=\varepsilon \Psi$ and therefore $y^{q}= \varepsilon^{q}  \Phi^{q}$ and $z^{r}= \varepsilon^{r}  \Psi^{r}$ for two positive integers $\Phi$ and $\Psi$. Therefore, in view of our hypothesis $x^{p} \pm y^{q}=z^{r}$, we should have 

\[x^{p}=\varepsilon^{r}  \Psi^{r} - \varepsilon^{q} \Phi^{q} = 
\left\{\begin{array}{ll}
 \ \varepsilon^{q} \left( \varepsilon^{r-q}  \Psi^{r} -  \Phi^{q} \right),\hspace{0.2cm} if \hspace{0.2cm} r\geq q\\
 \ \varepsilon^{r} \left( \Psi^{r} - \varepsilon^{q-r}  \Phi^{q} \right),\hspace{0.2cm} if \hspace{0.2cm} r\leq q. 
\end{array}\right.
\]
In particular, we should  get $ \varepsilon \mid x $, which violates directly our hypothesis that $gcd \left( x, y, z\right) =1$.\\
\textbf{ii}. To get a contradiction, let's assume that $gcd \left( y^{k}, z^{l}\right) = \epsilon >1$. Suppose $\delta$ is a prime factor of $\epsilon$. Since the integer $\delta$ is prime,it will necessarily divide both $y$ and $z$, i.e. $\delta \mid y$ and $\delta \mid z$. Obviously, from the relation $gcd \left( y^{k}, z^{l}\right) = \epsilon >1$ and the fact that $\delta$ is a prime factor of $\epsilon$, it follows directly that $y^{k}=\delta \tilde{\Phi} $ and $ z^{l}=\delta \tilde{\Psi}$ and, therefore, $y^{q}= \delta \tilde{\Phi} y^{q-k}$ and $z^{r}= \delta \tilde{\Psi} z^{r-l}$ for two positive integers $\tilde{\Phi}$ and $\tilde{\Psi}$. So, in view of our hypothesis $x^{p} + y^{q}=z^{r}$, we should  take 
$x^{p}=\delta \tilde{\Phi} y^{q-k} - \delta \tilde{\Psi} z^{r-l}  = \delta ( \tilde{\Phi} y^{q-k} - \tilde{\Psi} z^{r-l})$.
In particular, we should  get $ \delta \mid x^{p} $, which again violates our hypothesis that $gcd \left( x, y, z\right) =1$, because if a prime number, in this case the number $\delta$, divides a power of an integer then it will definitely divide that integer.\\
\noindent \textbf{iii}. To demonstrate this claim, it is sufficient to apply part \textit{ii} for $k=1$ and $l=2$ and obtain the following equations:\\
\indent $ 1=gcd \left( y, z^{2} \right) = gcd \left( y, z^{2} + \left( -\tau_{0}  y \right) y \right)$ \\ 
\indent \hspace{2.7cm} $ = gcd \left( - y^{\sigma q-2}, \left[ z^{2} - \tau_{0} y^{2} \right] \right)$. $\square$
\vspace{3mm}
\begin{proposition}  
With the notation adopted in Lemma 1 of the main text, the selected prime number $\pi$ fulfils:  
\begin{center}
$ y^{\sigma q-2} \left( m_{*}^{2}-\nu^{2} \right) =  [z^{2} - \tau_{0} y^{2} ] \mp \pi$.  
\end{center} 
for three suitably chosen positive integers $k$, $m_{*}$ and $\nu$.
\end{proposition}
\textit{Proof}. By \textit{Proposition 1 iii}, we have $gcd\left( - y^{\sigma q-2},\left[ z^{2} - \tau_{0} y^{2} \right] \right) =1$, so an application of Dirichlet’s Theorem on arithmetic progressions, in its basic form, shows that there are infinitely many positive integers $X$ such that all quantities $\pi=- y^{\sigma q-2} X + \left[ z^{2} - \tau_{0} y^{2} \right] $ represent prime numbers. For such a prime, we have successively:
\[ \pi= - y^{\sigma q-2} X + \left[ z^{2} -\tau_{0}  y^{2} \right]\Leftrightarrow 
 y^{\sigma q-2} X =  \left[  z^{2} - \tau_{0}  y^{2} \right] - \pi.
\]
Now, observe that, as already mentioned in \textit{Lemma 1}, since the numbers $\tau_{0}$, $y$ and $z$ are odd, the integer $X$ should also be an odd number and, therefore, it could be represented as the difference of the two squares $\left( \left[ X+1\right] /2 \right)^{2}$ and $\left( \left[ X-1\right] /2 \right)^{2}$. Consequently, the last equality we have reached above can equivalently be expressed in the form: \\
\indent \hspace{0.6cm} $ y^{\sigma q-2} \left( \underbrace{  \left( \left[ X+1\right] /2 \right)^{2}}_{m_{*}^{2} }   - \underbrace{\left( \left[ X-1\right] /2 \right)^{2}}_{ \nu^{2} } \right) = \left[ z^{2} - \tau_{0} y^{2} \right] - \pi $,\\
for two suitably chosen positive integers $ m_{*} $ and $ \nu $. Hence, it was proved that
\begin{center}
$y^{\sigma q-2} \left\lbrace  m_{*}^{2} -  \nu^{2}\right\rbrace  = \left[ z^{2} - \tau_{0} y^{2} \right] - \pi$,  
\end{center}
and the proof of the \textit{Proposition} is complete. $ \square $\\

\begin{proposition}  
If $  m_{*} $, $ \nu $ and  $\pi$ are as in the preceding Proposition, then the integers $n_{*}=-m_{*} \pm \nu $ are the two roots of the equation 
\begin{center}
$ y^{\sigma q-2}  n_{*}^{2}+  2 m_{*} y^{\sigma q-2}   n_{*} +\left( \left[ z^{2}- \tau_{0} y^{2} \right] - \pi \right) = 0$.  
\end{center} 
\end{proposition}
\textit{Proof}. Obviously, the  roots $ n_{*}$ of the quadratic polynomial $ \left[ y^{\sigma q-2} \right] n^{2}+ [  2 m_{*} y^{\sigma q-2}]$ $ n +\left( \left[ z^{2}- \tau_{0} y^{2} \right] - \pi \right)$ are given by
\begin{center}
$ n_{*}=- m_{*} \pm y^{2-\sigma q} \left\lbrace y^{\sigma q -2} \left[  m_{*}^{2}  y^{\sigma q -2}- \left( \left[ z^{2} - \tau_{0}  y^{2} \right] - \pi \right) \right] \right\rbrace ^{1/2}$. 
\end{center}
Since, by the preceding \textit{Proposition}, $ y^{\sigma q-2} \left\lbrace m_{*}^{2} -  \nu^{2}\right\rbrace  = \left[ z^{2} - \tau_{0}  y^{2} \right] - \pi$,  we can write $ \left[   m_{*}^{2}  y^{\sigma q -2}  - \left( \left[ z^{2} - \tau_{0} y^{2} \right] - \pi \right)\right]  = \left( \nu^{2}  y^{\sigma q -2} \right) $, and hence, the roots $n_{*}$ will necessarily be two integer numbers of the form $ n_{*}= -m_{*} \pm \nu $, which completes the proof. $ \square $\\

\begin{rem}
{\normalfont For the above mentioned options of $m_{*}$, $\nu$ and $n_{*}$, the two equations bellow coincide:

\indent $  y^{\sigma q-2} n_{*}^{2}+ 2 m_{*} y^{\sigma q-2}  n_{*} +\left( \left[ z^{2}- \tau_{0}  y^{2} \right] \mp \pi \right) = 0$\\
and\\
\indent $  y^{\sigma q-2} \left\lbrace m_{*}^{2} -  \nu^{2}\right\rbrace  = \left[ z^{2} - \tau_{0} y^{2} \right] - \pi$.} $\square $
\end{rem}
\vspace{0.3mm}

\begin{proposition}  
If $  m_{*} $, $n_{*}$ and  $\pi$ are as in Proposition 3, then\\

\indent \hspace{0.5cm}
$  y^{\sigma q-2} n_{*}^{2}+ 2 m_{*}  y^{\sigma q-2}   n_{*} +  \left[ z^{2} - \tau_{0} y^{2} \right] =  z^{2} -  y^{2} \tau_{m_{*}, n_{*}}=\pi $.
\end{proposition}
\textit{Proof}. Indeed, we have:\\

\indent $ z^{2} - y^{2} \tau_{m_{*}, n_{*}}= z^{2} -  y^{2} \left(-y^{\sigma q-4} n_{*}\left( 2m_{*}+n_{*} \right)  +\tau_{0} \right) $ \\ 
\indent \hspace{2.65cm} $ = z^{2} - \left( - y^{\sigma q-2} n_{*}\left( 2m_{*}+n_{*} \right)  +\tau_{0} y^{2}\right)$\\
\indent \hspace{2.65cm} $ = z^{2} +   y^{\sigma q-2} n_{*}\left( 2m_{*}+n_{*} \right)  - \tau_{0} y^{2}$\\
\indent \hspace{2.65cm} $ =\left(  y^{\sigma q-2}\right)  n_{*}^{2}+ \left( 2 m_{*}  y^{\sigma q-2} \right) n_{*} + \left[  z^{2}- \tau_{0}  y^{2} \right]$\\
\indent \hspace{2.65cm} $ =\pi$, by \textit{Proposition 3}.  $ \square $\\

\begin{proposition}  
If $U, V, \tilde{U},\tilde{V}\in \mathbb{Z}\setminus\lbrace0\rbrace$, the determinant of the matrix\\
$$
\begin{pmatrix}
U&-V&\hspace{0.1cm}\tau_{m_{*},n_{*}}V\\
\tilde{U}&-\tilde{V}&\hspace{0.1cm}\tau_{m_{*},n_{*}}\tilde{V}\\
 x^{2}& \hspace{0.2cm} y^{2} &\hspace{0.2cm}- z^{2}\\
\end{pmatrix}
$$
equals \\

\indent \hspace{3cm} $ \left( U\tilde{V}-\tilde{U}V \right) \left[ z^{2} -  y^{2} \tau_{m_{*}, n_{*}}\right]$.
\end{proposition}
\textit{Proof}. A straightforward computation gives:
\[
  det \begin{pmatrix}
U&-V&\hspace{0.1cm}\tau_{m_{*},n_{*}}V\\
\tilde{U}&-\tilde{V}&\hspace{0.1cm}\tau_{m_{*},n_{*}}\tilde{V}\\
 x^{2}& \hspace{0.2cm} y^{2} &\hspace{0.2cm}- z^{2}\\
\end{pmatrix}=U \tilde{V}  z^{2} - U \tilde{V}  \tau_{m_{*},n_{*}} y^{2} - \tilde{U} V z^{2}    
\]
\indent \hspace{4.2cm} $ - V \tilde{V}  \tau_{m_{*},n_{*}} x^{2} + \tilde{U} V \tau_{m_{*},n_{*}} y^{2} +  V \tilde{V} \tau_{m_{*},n_{*}} x^{2}$\\
\indent \hspace{3cm} $ =U \tilde{V} z^{2}  + \tilde{U} V \tau_{m_{*},n_{*}} y^{2} - U \tilde{V} \tau_{m_{*},n_{*}} y^{2} - \tilde{U} V  z^{2}  $\\
\indent \hspace{3cm} $ = \left( U\tilde{V}-\tilde{U}V \right)  z^{2} - \left( U\tilde{V}-\tilde{U}V \right)  y^{2} \tau_{m_{*}, n_{*}} $\\
\indent \hspace{3cm} $ = \left( U\tilde{V}-\tilde{U}V \right) \left[  z^{2} -  y^{2} \tau_{m_{*}, n_{*}}\right]$. $\square $

\begin{proposition}  
The system in the Proof of Theorem 1\\
\[\left( \mathbb{S}_{\mathbb{Ζ}}\right) 
\left\{\begin{array}{ll}
 \ (U)X+ (-V)Y+(\tau_{m_{*},n_{*}}V)Z= 1\\
 \ (\tilde{U})X+ (-\tilde{V})Y+(\tau_{m_{*},n_{*}}\tilde{V})Z= 1\\
 \ ( x^{2})X + ( y^{2})Y+(-z^{2})Z=0, 
\end{array}\right.
\]
with determinant $\left( U\tilde{V}-\tilde{U}V\right)\left[ z^{2} - y^{2}\tau_{m_{*},n_{*}}\right]\neq0$ {\normalfont( see} Proposition 5 {\normalfont above)}, has unique solution\\

$X= - \frac{V-\tilde{V}}{ U\tilde{V}-\tilde{U}V}$, \\

$Y= - \frac{\left( U-\tilde{U}\right) z^{2}+\left( V-\tilde{V}\right)\tau_{m_{*},n_{*}} x^{2}}{\left( U\tilde{V}-\tilde{U}V\right)\left[ z^{2} -  y^{2}\tau_{m_{*},n_{*}}\right]}$ and \\

$Z= -\frac{\left( U-\tilde{U}\right) y^{2} + \left( V-\tilde{V}\right) x^{2}}{\left( U\tilde{V}-\tilde{U}V\right)\left[ z^{2} -  y^{2}\tau_{m_{*},n_{*}}\right]}$. 
\end{proposition}
\textit{Proof}. We have\\
$ X=\frac{det
  \left( {\begin{array}{ccc}
   1&-V&\hspace{0.1cm}\tau_{m_{*},n_{*}}V\\
   1&-\tilde{V}&\hspace{0.1cm}\tau_{m_{*},n_{*}}\tilde{V}\\
   0& \hspace{0.2cm} y^{2} &\hspace{0.2cm}-z^{2}\\
  \end{array} } \right)}{\left( U\tilde{V}-\tilde{U}V\right)\left[ z^{2} -  y^{2}\tau_{m_{*},n_{*}}\right]} $ 
$= \frac{  \tilde{V} z^{2}-\tau_{m_{*},n_{*}} \tilde{V}  y^{2}- V z^{2}+\tau_{m_{*},n_{*}} V y^{2}}{\left( U\tilde{V}-\tilde{U} V \right)\left[ z^{2} -  y^{2}\tau_{m_{*},n_{*}}\right]}$ \\

\indent \hspace{1,5cm} $= \frac{ \left( - V + \tilde{V} \right) z^{2} + \left( V - \tilde{V} \right) \tau_{m_{*},n_{*}}  y^{2}}{\left( U\tilde{V}-\tilde{U} V \right)\left[ z^{2} -  y^{2}\tau_{m_{*},n_{*}}\right]} = \frac{ \left( V - \tilde{V} \right) \left[ - z^{2}\right]  + \left( V - \tilde{V} \right) \tau_{m_{*},n_{*}}  y^{2}}{\left( U\tilde{V}-\tilde{U} V \right)\left[ z^{2} -  y^{2}\tau_{m_{*},n_{*}}\right]}$  \\

\indent \hspace{1,5cm} $= \frac{ \left( V - \tilde{V} \right) \left( - z^{2}+  \tau_{m_{*},n_{*}}  y^{2} \right) }{\left( U\tilde{V}-\tilde{U} V \right)\left[ z^{2} -  y^{2}\tau_{m_{*},n_{*}}\right]} = - \frac{ \left( V - \tilde{V} \right) \left( z^{2}- \tau_{m_{*},n_{*}}  y^{2} \right) }{\left( U\tilde{V}-\tilde{U} V \right)\left[ z^{2} -  y^{2}\tau_{m_{*},n_{*}}\right]}$ $=- \frac{V-\tilde{V}}{ U\tilde{V}-\tilde{U}V} $,\\
\\

$ Y=\frac{det
  \left( {\begin{array}{ccc}
   U& 1&\hspace{0.1cm}\tau_{m_{*},n_{*}}V\\
   \tilde{U}& 1&\hspace{0.1cm}\tau_{m_{*},n_{*}}\tilde{V}\\
   x^{2}& \hspace{0.2cm} 0 &\hspace{0.2cm}-z^{2}\\
  \end{array} } \right)}{\left( U\tilde{V}-\tilde{U}V\right)\left[ z^{2} -  y^{2}\tau_{m_{*},n_{*}}\right]} $ 
$= \frac{ \left( - U + \tilde{U} \right) z^{2} + \left( - V + \tilde{V} \right) \tau_{m_{*},n_{*}}  x^{2}}{\left( U\tilde{V}-\tilde{U} V \right)\left[ z^{2} -  y^{2}\tau_{m_{*},n_{*}}\right]} = \frac{ - \left( U - \tilde{U} \right) z^{2}  - \left( V - \tilde{V} \right) \tau_{m_{*},n_{*}}  x^{2}}{\left( U\tilde{V}-\tilde{U} V \right)\left[ z^{2} -  y^{2}\tau_{m_{*},n_{*}}\right]}$\\

\noindent and\\

$ Z=\frac{det
  \left( {\begin{array}{ccc}
   U&-V& 1\\
   \tilde{U}&-\tilde{V}& 1\\
   x^{2}&  y^{2}&0\\
  \end{array} } \right)}{\left( U\tilde{V}-\tilde{U}V\right)\left[ z^{2} -  y^{2}\tau_{m_{*},n_{*}}\right]} $ 
 $= \frac{ \left( - U + \tilde{U} \right) y^{2} + \left( - V + \tilde{V} \right) x^{2}}{\left( U\tilde{V}-\tilde{U} V \right)\left[ z^{2} -  y^{2}\tau_{m_{*},n_{*}}\right]} = \frac{ - \left( U - \tilde{U} \right)y^{2}  - \left( V - \tilde{V} \right)  x^{2}}{\left( U\tilde{V}-\tilde{U} V \right)\left[ z^{2} - y^{2}\tau_{m_{*},n_{*}}\right]}. \square $\\

\begin{proposition}  
Given any partial integer solution $\left( u, v \right) \in \mathbb{Z}^{2}$ of the Diophantine equation $x^{p-2} u+\left(\tau_{m_{*},n_{*}}z^{r-2} -y^{q-2}\right)v= 1$, the integer numbers\\
\indent $ U:=u+\kappa \left( \tau_{m_{*},n_{*}} z^{r-2}-y^{q-2} \right)$,\\
\indent $V:=v-\kappa x^{p-2}$,\\
\indent $\tilde{U}:=u+\ell\left( \tau_{m_{*},n_{*} } z^{r-2}-y^{q-2} \right) $ and\\
\indent $\tilde{V}:=v-\ell x^{p-2}$\\
satisfy \\
\indent \hspace{3,5cm} $U, V, \tilde{U}, \tilde{V} \neq 0$\\
\noindent whenever $\kappa \in \mathbb{Ζ}^{+}$, $\ell \in \mathbb{Ζ}^{+}$, $\kappa\neq \ell $.
\end{proposition}
\textit{Proof}. Indeed, it is easily verified that
\begin{itemize}
\item If $U=0 \Leftrightarrow u+\kappa \left( \tau_{m_{*},n_{*}} z^{r-2}-y^{q-2} \right) =0$, then, by equation $x^{p-2}u+\left(\tau_{m_{*},n_{*}}z^{r-2}  -  y^{q-2} \right)v= 1$, we should have $-\kappa x^{p-2} \left( \tau_{m_{*},n_{*}} z^{r-2}-y^{q-2}\right) + \left( \tau_{m_{*},n_{*}} z^{r-2}-y^{q-2}\right)v = 1$  $ \Leftrightarrow $  $ \left( v - \kappa x^{p-2} \right) \left( \tau_{m_{*},n_{*}} z^{r-2}-y^{q-2}\right) = 1$. Or, equivalently,   $ V \left( \tau_{m_{*},n_{*}} z^{r-2} -y^{q-2} \right) = 1$, which is absurd in $\mathbb{Z}$, since $V$ is an integer and, by \textit{Lemma 2}, $\tau_{m_{*},n_{*}} z^{r-2} -y^{q-2} \neq 1$.
  \item In like manner, if $\tilde{U}=0 \Leftrightarrow u+\ell \left( \tau_{m_{*},n_{*}} z^{r-2}-y^{q-2} \right) =0$, then, again by the same as above equation (i.e., $x^{p-2}u+\left(\tau_{m_{*},n_{*}}z^{r-2}-y^{q-2} \right)v= 1$), we should have $-\ell x^{p-2} \left(  \tau_{m_{*},n_{*}} z^{r-2} - y^{q-2} \right)+ \left( \tau_{m_{*},n_{*}} z^{r-2} -  y^{q-2}\right)v = 1 \Leftrightarrow \left( v - \ell x^{p-2} \right) \left( \tau_{m_{*},n_{*}}  z^{r-2}-y^{q-2}\right) = 1 \Leftrightarrow \tilde{V} \left( \tau_{m_{*},n_{*}} z^{r-2} -y^{q-2}\right) = 1$ which is absurd in $\mathbb{Z}$, since $\tilde{V}$ is an integer and, again by \textit{Lemma 2}, $\tau_{m_{*},n_{*}} z^{r-2} -y^{q-2} \neq 1$.
  \item On the other hand, if $V=0 \Leftrightarrow v-\kappa x^{p-2}=0$, then once more from the equation $x^{p-2} u+\left(\tau_{m_{*},n_{*}}z^{r-2} - y^{q-2} \right)v= 1$, we should come to the conclusion $x^{p-2} \left( u + \kappa \left[ \tau_{m_{*},n_{*}} z^{r-2} -y^{q-2} \right] \right) = 1$ which is absurd in $\mathbb{Z}$, since $x^{p-2}$ and $u+\kappa \left[ \tau_{m_{*},n_{*}} z^{r-2}-y^{q-2} \right]$ are integers and $x^{p-2}>1$.
   \item Similarly, if $\tilde{V}=0 \Leftrightarrow v-\ell x^{p-2}=0$, then from equation $x^{p-2} u+(  \tau_{m_{*},n_{*}}z^{r-2} - y^{q-2}) v= 1$ it would follow that $ x^{p-2} u+ \left(  \tau_{m_{*},n_{*}} z^{r-2}- y^{q-2} \right) \ell x^{p-2} = 1  \Leftrightarrow x^{p-2} \left( u + \ell \left[ \tau_{m_{*},n_{*}} z^{r-2} -y^{q-2} \right] \right) = 1$ which is absurd in $\mathbb{Z}$, since $x^{p-2}$ and $u+\ell \left[ \tau_{m_{*},n_{*}} z^{r-2}-y^{q-2} \right]$ are integers and $x^{p-2}>1$.$\square$
\end{itemize} 

\begin{proposition}  
If $ U:=u+\kappa \left( \tau_{m_{*},n_{*}} z^{r-2}-y^{q-2} \right)$, $V:=v-\kappa x^{p-2}$, $\tilde{U}:=u+\ell\left( \tau_{m_{*},n_{*} } z^{r-2}-y^{q-2} \right) $, $\tilde{V}:=v-\ell x^{p-2}$ and the parameters $u$ and $v$ satisfy the equation $x^{p-2} u+\left(\tau_{m_{*},n_{*}}z^{r-2} -y^{q-2}\right)v= 1$, then \\
\indent \hspace{3,5cm} $\left( U\tilde{V}-\tilde{U}V\right)= \kappa - \ell $.
\end{proposition}
\textit{Proof}.	Indeed, we have
\begin{equation*}
\left.\begin{aligned}
  U&=u+\kappa \left( \tau_{m_{*},n_{*}} z^{r-2}-y^{q-2} \right)\\
  \tilde{V}&=v-\ell x^{p-2}
\end{aligned}\right\} 
\Rightarrow  U\tilde{V}=
\left\{\begin{array}{ll}
 \ uv-\ell u x^{p-2}\\
 \ \hspace{0,1cm} +v \kappa \left( \tau_{m_{*},n_{*}} z^{r-2} - y^{q-2} \right)\\
 \ \hspace{0,2cm} -\kappa \ell\left(\tau_{m_{*},n_{*}} z^{r-2} - y^{q-2} \right) x^{p-2}, 
\end{array}\right.
\end{equation*}
\indent \hspace{4,8cm} $  -  $
\begin{equation*}
\left.\begin{aligned}
 \tilde{U}&=u+\ell \left( \tau_{m_{*},n_{*}} z^{r-2}-y^{q-2} \right)\\
  V &=v-\kappa x^{p-2}
\end{aligned}\right\} 
\Rightarrow  \tilde{U} V=
\left\{\begin{array}{ll}
 \ uv-\kappa u x^{p-2}\\
 \ \hspace{0,1cm} +v \ell \left( \tau_{m_{*},n_{*}} z^{r-2} - y^{q-2} \right)\\
 \ \hspace{0,2cm} -\kappa \ell\left(\tau_{m_{*},n_{*}} z^{r-2} - y^{q-2} \right) x^{p-2}, 
\end{array}\right.
\end{equation*}
\indent \hspace{4,8cm} $  \Downarrow  $\\

\noindent$  U\tilde{V}-\tilde{U} V=-\ell u x^{p-2}+ \kappa u x^{p-2} + v \kappa \left( \tau_{m_{*},n_{*}} z^{r-2} - y^{q-2} \right)-v \ell \left( \tau_{m_{*},n_{*}} z^{r-2} - y^{q-2} \right)$\\

\indent \hspace{0,46cm} $  = \left( \kappa - \ell \right)u x^{p-2} + \left( \kappa - \ell \right)v \left( \tau_{m_{*},n_{*}} z^{r-2} - y^{q-2} \right)$\\

\indent \hspace{0,46cm} $  = \left( \kappa - \ell \right) ( \underbrace{  u x^{p-2} +v \left( \tau_{m_{*},n_{*}} z^{r-2} - y^{q-2} \right) }_{= 1}) $\\
\indent \hspace{0,46cm} $  = \kappa - \ell $. $ \square $\\

\begin{proposition}
The unique solution $(X, Y, Z)$ of the system $\left( \mathbb{S}_{\mathbb{Z}}\right)$ in the Proof of Theorem 1 may be expressed in the form
\begin{center} 
$X=x^{p-2}$, $Y=\left( - z^{2} \right) D + \left( x^{p} \tau_{m_{*},n_{*}} \right) E$ and $Z=\left( - y^{2} \right) D+\left( x^{p} \right) E x^{p}$,
\end{center}
where 
\begin{center}
$D:=\frac{\tau_{m_{*},n_{*}}z^{r-2} - y^{q-2}} {z^{2} - y^{2}\tau_{m_{*},n_{*}}}$ and $E:=\frac{1}{z^{2} - y^{2}\tau_{m_{*},n_{*}}}$.  
\end{center} 
With this formulation, the following system of equations applies\\
\[ 
\left\{\begin{array}{ll}
 \ y^{q-2} =\, \left( - z^{2} \right) D +\left(  x^{p} \tau_{m_{*},n_{*}} \right) E\\
 \ z^{r-2} =\, \left( - y^{2} \right) D +\left(  x^{p} \right) E.
\end{array}\right.
\]
\end{proposition}
\textit{Proof}.	By performing standard calculations, we successively obtain:\\

$X= -\frac{V-\tilde{V}}{U\tilde{V}-\tilde{U}V}=
- \frac{-\kappa x^{p-2}+\ell x^{p-2}}{ \kappa - \ell }=  -\frac{-\kappa +\ell}{ \kappa - \ell } x^{p-2} =x^{p-2} $\\

\indent \hspace{0.5cm}$ \Rightarrow x^{p-2}=x^{p-2}$ (:trivial identity),\\

$Y = \frac{- \left( U-\tilde{U} \right) z^{2} - \left( V-\tilde{V} \right)\tau_{m_{*},n_{*}} x^{2}}{\left( U\tilde{V}-\tilde{U}V\right) \left( z^{2} - y^{2}\tau_{m_{*},n_{*}} \right) } $\\

\indent \hspace{0,40cm} $= \frac{- \left( u+\kappa \left( \tau_{m_{*},n_{*}} z^{r-2}-y^{q-2} \right) -u-\ell\left( \tau_{m_{*},n_{*}} z^{r-2}-y^{q-2} \right)\right) z^{2}- \left( v-\kappa x^{p-2}-v+\ell x^{p-2} \right)  \tau_{m_{*},n_{*}} x^{2}}{\left( \kappa - \ell \right)\left( z^{2} - y^{2}\tau_{m_{*},n_{*}} \right)} $\\

\indent \hspace{0,40cm} $ = \frac{- \left( \kappa -\ell \right) \left( \tau_{m_{*},n_{*}} z^{r-2}-y^{q-2} \right) z^{2} - \left( -\kappa +\ell \right) \tau_{m_{*},n_{*}} x^{2}x^{p-2}}{ \left( \kappa - \ell \right)\left( z^{2} - y^{2}\tau_{m_{*},n_{*}} \right)} $\\

\indent \hspace{0,40cm} $ = \frac{ - \left( \kappa -\ell \right) \left( \tau_{m_{*},n_{*}} z^{r-2}-y^{q-2} \right) z^{2}}{\left( \kappa - \ell \right)\left( z^{2} - y^{2}\tau_{m_{*},n_{*}} \right)}+\frac{ - \left( -\kappa +\ell \right) \tau_{m_{*},n_{*}} x^{2}x^{p-2}}{ \left( \kappa - \ell \right)\left( z^{2} - y^{2}\tau_{m_{*},n_{*}} \right)} $\\

\indent \hspace{0,40cm} $= -\frac{ \left( \tau_{m_{*},n_{*}} z^{r-2}-y^{q-2} \right) z^{2}}{ z^{2} - y^{2}\tau_{m_{*},n_{*}} }+\frac{\tau_{m_{*},n_{*}} x^{p}}{ z^{2} - y^{2}\tau_{m_{*},n_{*}} } $\\

$\Rightarrow Y=  -\underbrace{\frac{\tau_{m_{*},n_{*}} z^{r-2}-y^{q-2}}{ z^{2} - y^{2}\tau_{m_{*},n_{*}} }}_{=D} z^{2} 
      +\underbrace{\frac{1}{ z^{2} - y^{2}\tau_{m_{*},n_{*}}}}_{=E} \tau_{m_{*},n_{*}} x^{p} $\\

\indent \hspace{0.5cm} $ \Rightarrow y^{q-2}= \left( - z^{2} \right) D +\left(  x^{p} \tau_{m_{*},n_{*}} \right) E$

\noindent and 

$Z= \frac{- \left( U-\tilde{U} \right) y^{2} - \left( V-\tilde{V} \right) x^{2}}{\left( U\tilde{V}-\tilde{U}V\right) \left( z^{2} - y^{2}\tau_{m_{*},n_{*}} \right) }$\\

\indent \hspace{0,40cm} $=  \frac{- \left( u+\kappa \left( \tau_{m_{*},n_{*}} z^{r-2}-y^{q-2} \right) -u-\ell\left( \tau_{m_{*},n_{*}} z^{r-2}-y^{q-2} \right)\right)  y^{2}- \left( v-\kappa x^{p-2}-v+\ell x^{p-2} \right) x^{2}}{\left( \kappa - \ell \right)\left( z^{2} - y^{2}\tau_{m_{*},n_{*}} \right)} $\\

\indent \hspace{0,40cm} $= \frac{- \left( \kappa -\ell \right) \left( \tau_{m_{*},n_{*}} z^{r-2}-y^{q-2} \right) y^{2}- \left( -\kappa +\ell \right) x^{2}x^{p-2}}{\left( \kappa - \ell \right)\left( z^{2} - y^{2}\tau_{m_{*},n_{*}} \right)}  $\\

\indent \hspace{0,40cm} $= \frac{- \left( \kappa -\ell \right) \left( \tau_{m_{*},n_{*}} z^{r-2}-y^{q-2} \right)  y^{2}}{ \left( \kappa - \ell \right)\left( z^{2} - y^{2}\tau_{m_{*},n_{*}} \right)}+\frac{- \left( -\kappa +\ell \right)  x^{p}}{ \left( \kappa - \ell \right)\left( z^{2} - y^{2}\tau_{m_{*},n_{*}} \right)}  $\\

\indent \hspace{0,40cm} $=  -\frac{ \left( \tau_{m_{*},n_{*}} z^{r-2}-y^{q-2} \right) y^{2}}{ z^{2} - y^{2}\tau_{m_{*},n_{*}} }+\frac{ x^{p}}{ z^{2} - y^{2}\tau_{m_{*},n_{*}} }  $\\

$\Rightarrow Z= - \underbrace{\frac{\tau_{m_{*},n_{*}} z^{r-2}-y^{q-2}}{ z^{2} - y^{2}\tau_{m_{*},n_{*}} }}_{=D} y^{2} 
      +\underbrace{\frac{1}{ z^{2} - y^{2}\tau_{m_{*},n_{*}}}}_{=E} x^{p}  $\\

\indent \hspace{0.5cm} $ \Rightarrow z^{r-2} = \left( -  y^{2} \right) D + \left(  x^{p} \right) E$. $\square$\\

\begin{proposition}\textbf{i.}
If $x^{p}+ y^{q}=z^{r}$, the system of Proposition 9 is written as

\[\left( \mathbb{S} \right)
 \left\{\begin{aligned}
   \left( 1-Ez^{2} \right) Z + \left( Ey^{2} \right) Y   &= -  D y^{2} \\
   \left( \tau_{m_{*},n_{*}} Ez^{2} \right)Z+\left( -Ey^{2}\tau_{m_{*},n_{*}}- 1\right) Y  &=   D z^{2}       
  \end{aligned}\right.
\]
where we have used the notation $ Z=z^{r-2}$ and $Y =y^{q-2} $.\\
\textbf{ii.} The system $ \mathbb{S}$ has no integer solutions $Z$, $Y$.
\end{proposition}
\textit{Proof}.	\textbf{i.} The system of \textit{Proposition 9} can be successively written as follows

\[ 
\left\{
\begin{array}{lll}
      
       &  \left( -z^{2} \right) D +\left(  x^{p} \tau_{m_{*},n_{*}} \right) E &= y^{q-2}\\
      
       & \left( -  y^{2} \right) D +\left(  x^{p} \right) E &=  z^{r-2}\\
\end{array} 
\right. 
\]

\[ 
\indent \hspace{0.2cm}    \Rightarrow \left\{
\begin{array}{lll}
      
       & \left( - z^{2} \right) D + \tau_{m_{*},n_{*}}  E z^{r} - \tau_{m_{*},n_{*}}  E y^{q} &= y^{q-2}\\
      
       & \left( -  y^{2} \right) D + E z^{r} -  E y^{q} &=  z^{r-2} \\
\end{array} 
\right. 
\]

\[ 
\indent \hspace{0.2cm} \Rightarrow \left\{
\begin{array}{lll}
      
      & \left( \tau_{m_{*},n_{*}}  E z^{2} \right) z^{r-2} +\left( - \tau_{m_{*},n_{*}}  E y^{2} -1 \right) y^{q-2} &= D z^{2}\\
      
      &  \left( 1-  E z^{2} \right) z^{r-2} + \left( E y^{2} \right) y^{q-2} &=  -Dy^{2} \\     
\end{array} 
\right. 
\]
   
\[ 
\indent \hspace{0.2cm} \Rightarrow \left\{
\begin{array}{lll}
      
     \left( \tau_{m_{*},n_{*}}  E z^{2} \right) z^{r-2} +\left( - \tau_{m_{*},n_{*}}  E y^{2} - 1 \right) y^{q-2} &= D z^{2}\\
      
     \left( 1-  E z^{2} \right) z^{r-2} + \left( E y^{2} \right) y^{q-2} &=  - Dy^{2}. \\

\end{array} 
\right. 
\]
   
\noindent \textbf{ii.}  Let us now consider the system:

\[\left( \mathbb{S^{'}} \right)
 \left\{\begin{aligned}
   \left( \tau_{m_{*},n_{*}} E z^{2} \right) Z +\left( - \tau_{m_{*},n_{*}}  E y^{2} - 1 \right) Y  &=  D z^{2} \\
   \left( 1-  E z^{2} \right) Z + \left( E y^{2} \right) Y    &= - Dy^{2}.      
  \end{aligned}\right.
\]

As it just turned out, this system has the solution $\left( Z,Y\right) =\left( z^{r-2}, y^{q-2}\right) $. However, the determinant of this system equals\\

\indent \hspace{0.18cm} $det \left( \mathbb{S^{'}} \right)=  \left( Ey^{2} \right) \left( \tau_{m_{*},n_{*}} Ez^{2} \right) - \left( 1-Ez^{2} \right) \left( -\tau_{m_{*},n_{*}} E y^{2} - 1 \right) $ \\
\indent \hspace{1.5cm} $= \tau_{m_{*}, n_{*}} E^{2} y^{2} z^{2} + \tau_{m_{*}, n_{*}} E y^{2}+1 - \tau_{m_{*}, n_{*} }  E^{2} y^{2} z^{2}- E z^{2} $\\
\indent \hspace{1.5cm} $= \tau_{m_{*}, n_{*}} E y^{2} + 1- E z^{2}$ \\
\indent \hspace{1.5cm} $= 1- \left( z^{2}- \tau_{m_{*},n_{*}} y^{2} \right) E = 0$, by the definition of $E$.\\
 
This means that the system $( \mathbb{S^{'}})$  has no solution or has infinitely many solutions. Given that it is not possible for the system $\mathbb{S^{'}}$ to have no solution, since the pair $(Y, Z)=\left(y^{q-2}, z^{r-2} \right)$ is a pair of integer solutions of $( \mathbb{S^{'}} )$, we conclude that this system should have an infinite number of solutions. But, this directly contradicts our construction, since in such a case, we would have ($\tau_{m^{*}, n^{*}} E z^{2}=C(1-Ez^{2})$, $(-E \tau_{m^{*}, n^{*}} y^{2}-1)= C Ey^{2}$ and especially) \\

\indent \hspace{2cm} $D z^{2}=-CD y^{2}$, for a $C\in \mathbb{Z}$,\\
  
\noindent so we would infer $z^{2}=-C y^{2}$. But, if $z^{2}=-C y^{2}$, then the prime number $z^{2}-y^{2}\tau_{m^{*}, n^{*}} $ (see \textit{Lemma 1}) would be expressed as a product of two integers:\\

\indent \hspace{1cm} $z^{2}-y^{2}\tau_{m^{*}, n^{*}} =-Cy^{2}-y^{2}\tau_{m^{*}, n^{*}}=y^{2} \left( -C-\tau_{m^{*}, n^{*}} \right) $, \\

\noindent which is impossible to happen. $\square$\\

\vspace{10mm}

\begin{large}
\indent \hspace{4.5cm}\textbf{Appendix B}
\end{large}%

\vspace{5mm}

As in the Proof of Theorem 2, let us consider the \textit{exponential sum}:
\begin{center}
$\mathcal{S}_{k}^{(\ell)}= \sum_{x=0}^{\mathcal{Ν}-1} e^{2\pi i \frac{k x^{\ell} }{\mathcal{Ν}}}= e^{2\pi i \frac{k }{\mathcal{Ν}}0^{\ell}} + e^{2\pi i \frac{k }{\mathcal{Ν}}1^{\ell}}+...+e^{2\pi i \frac{k }{\mathcal{Ν}}\left( \mathcal{Ν}-1\right) ^{\ell}}  $
\end{center}
whenever $\ell \in \mathbb{N}$, $\mathcal{Ν}$ is a prime number and $k\in \mathbb{Z}_{\mathcal{Ν}}=\lbrace 0, 1, 2,..., \mathcal{Ν}-1\rbrace$. Then\\

\indent \hspace{1.7cm} $\sum_{k \in \mathbb{Z}_{\mathcal{Ν}}} \mathcal{S}_{k}^{(p)} \conjug{\mathcal{S}_{k}^{(q)}}=
\sum_{x, y \in \mathbb{Z}_{\mathcal{Ν}}} \sum_{k=0}^{\mathcal{Ν}-1} e^{2 \pi i k \frac{x^{p}-y^{q}}{\mathcal{Ν}}}$.\\

\begin{lemm}
\indent \hspace{1.2cm} $\sum_{k=0}^{\mathcal{Ν}-1} e^{2 \pi i k \frac{x^{p}-y^{q}}{\mathcal{Ν}}}= 
\begin{cases}
1 & \text{if } x^{p} \equiv     y ^{q} (mod \mathcal{Ν})\\
0 & \text{if } x^{p} \not\equiv y^{q}  (mod \mathcal{Ν}).
\end{cases}$\\
\end{lemm}
\textit{Proof}. Indeed, we have\\

\indent \hspace{0.6cm} $\sum_{k=0}^{\mathcal{Ν}-1} e^{2 \pi i k \frac{x^{p}-y^{q}}{\mathcal{Ν}}}= 
\sum_{k=0}^{\mathcal{Ν}-1} \left(  e^{\frac{2 \pi i\left( x^{p}-y^{q}\right) }{\mathcal{Ν}}} \right)^{k} =\frac{\left( e^{\frac{2 \pi i \left( x^{p}-y^{q} \right)}{\mathcal{Ν}}}\right)^{(\mathcal{Ν}-1)+1} -1}
{e^{\frac{2 \pi i \left( x^{p}-y^{q} \right)}{\mathcal{Ν}}}-1}$\\

\indent \hspace{3.4cm} $=\frac{\left( e^{\frac{2 \pi i \left( x^{p}-y^{q} \right)}{\mathcal{Ν}}}\right)^{\mathcal{Ν}} -1}
{e^{\frac{2 \pi i \left( x^{p}-y^{q} \right)}{\mathcal{Ν}}}-1}=\frac{ e^{2 \pi i \left( x^{p}-y^{q} \right)} -1} {e^{\frac{2 \pi i \left( x^{p}-y^{q} \right)}{\mathcal{Ν}}}-1}$\\

\indent \hspace{3.4cm} $=\frac{\left(  e^{2 \pi i }\right)^{\left( x^{p}-y^{q} \right)} -1} {e^{\frac{2 \pi i \left( x^{p}-y^{q} \right)}{\mathcal{Ν}}}-1}=
\begin{cases}
1 & \text{if } x^{p} \equiv     y ^{q} (mod \mathcal{Ν})\\
0 & \text{if } x^{p} \not\equiv y^{q}  (mod \mathcal{Ν}).  \square
\end{cases}
$\\

\indent By Lemma 3, we deduce the next auxiliary result.

\begin{lemm} The following relationship applies
\begin{center}
$\sum_{k \in \mathbb{Z}_{\mathcal{Ν}}} \mathcal{S}_{k}^{(p)} \conjug{\mathcal{S}_{k}^{(q)}}=
\sum_{x, y \in \mathbb{Z}_{\mathcal{Ν}}} \sum_{k \in \mathbb{Z}_{\mathcal{Ν}}} e^{2 \pi i k \frac{x^{p}-y^{q}}{\mathcal{Ν}}}= 
\mathcal{Ν} \mathcal{A}_{p, q}$
\end{center}
where $\mathcal{A}_{p, q}=\mathcal{A}_{p, q}^{(\mathcal{Ν})}$ is the number of solutions to the equation $x^{p}=y^{q}$ within the field $\mathbb{Z}_{\mathcal{Ν}}$, i.e.\\
\indent \hspace{2.0cm} $ \mathcal{A}_{p, q}=\mathcal{A}_{p, q}^{(\mathcal{Ν})}= \mid \left\lbrace x,y \in \mathbb{Z}_{\mathcal{Ν}}: x^{p} \equiv     y ^{q} (mod \mathcal{Ν}) \right\rbrace \mid$. $\square$\\
\end{lemm}

Further, with the notation adopted in Lemma 3, we have the following upper bound for $\mathcal{A}_{p, q}=\mathcal{A}_{p, q}^{(\mathcal{Ν})}$:

\begin{lemm} If $p$ and $q$ are natural numbers that are greater than $1$ and such that \\

\indent \hspace{5cm} $q \mid p$ or $p \mid q$,\\

\noindent then\\
\indent \hspace{3.4cm} $ \mathcal{A}_{p, q} \leq 1+ min \left\lbrace  p, q \right\rbrace \left(\mathcal{Ν} -1 \right) $.
\end{lemm}
\textit{Proof}. Suppose $q \mid p$. When $x=0$, we have only one solution to the equation $x^{p} \equiv y ^{q} (mod \mathcal{Ν})$, namely when $y=0$. Now, if we fix $x\in \mathbb{Z}_{\mathcal{Ν}}$, we have at most $q-1$ solutions $y$ to the equation $x^{p} \equiv y ^{q} (mod \mathcal{Ν})$. Indeed, the idea is if $x\neq 0$, then writing $y=ux^{p/q}$, where \\
\indent \hspace{2.0cm}$u=cos\left( 2k \pi \frac{p}{q}\right) +i sin\left( 2k \pi \frac{p}{q}\right) $, for $k=0,1,..., q-1$. \\
Since there are $\mathcal{Ν}-1$ choices for $x$,  this means that\\
\indent \hspace{2.2cm} $ \mathcal{A}_{p, q} \leq 1+  q \left(\mathcal{Ν} -1 \right)= 1+ min \left\lbrace  p, q \right\rbrace \left(\mathcal{Ν} -1 \right) $.\\
If  $p \mid q$, we derive the same inequality, using the analogous reasoning. $\square$\\ 
\\

Now, for $p \in \mathbb{N}$, put $\mathbb{G}_{p}:= \left\lbrace a^{p}: a \in \mathbb{Z}_{\mathcal{Ν}}^{*} \right\rbrace$. Since $\mathcal{S}_{k}^{(p)}=\mathcal{S}_{ka^{p}}^{(p)}$ whenever $a \in \mathbb{Z}_{\mathcal{Ν}}^{*} $, we infer
\begin{equation}
\sum_{k \in \mathbb{Z}_{\mathcal{Ν}}^{*}} \lvert \mathcal{S}_{k}^{(p)} \rvert^{2} \geq 
\sum_{a^{p} \in \mathbb{G}_{p}} \lvert \mathcal{S}_{ka^{p}}^{(p)} \rvert^{2}= \lvert \mathbb{G}_{p}\rvert \times \lvert \mathcal{S}_{k}^{(p)} \rvert^{2}
\end{equation}
By Lemma 4, we have
\begin{center}
$\sum_{k \in \mathbb{Z}_{\mathcal{Ν}}^{*}} \mid \mathcal{S}_{k}^{(p)} \mid^{2} \leq \mathcal{Ν} \mathcal{A}_{p, q}$
\end{center}
while, by Lemma 5, 
\begin{center}
$ \mathcal{Ν} \mathcal{A}_{p, q} \leq \mathcal{Ν} \left( 1+ min \left\lbrace  p, q \right\rbrace \left(\mathcal{Ν} -1 \right) \right)$.
\end{center}
Since $ \mathcal{Ν} \left( 1+ min \left\lbrace  p, q \right\rbrace \left(\mathcal{Ν} -1 \right) \right) \leq p \mathcal{Ν}^{2}  $, we infer
\begin{center}
$\sum_{k \in \mathbb{Z}_{\mathcal{Ν}}} \lvert \mathcal{S}_{k}^{(p)} \rvert^{2} \leq p \mathcal{Ν}^{2} $.
\end{center}
In view of (10), this implies that
\begin{equation}
\lvert \mathcal{S}_{k}^{(p)} \rvert^{2} \leq \frac{p \mathcal{Ν}^{2}}{\lvert \mathbb{G}_{p} \rvert } .
\end{equation}

Next, observe that, given any $x\in \mathbb{Z}_{\mathcal{Ν}}^{*}$, we have $x^{p} \in \mathbb{G}_{p}$. We partition $\mathbb{Z}_{\mathcal{Ν}}^{*}$  into $\lvert \mathbb{G}_{p} \rvert$ sets:
\begin{center}
$ \left\lbrace x \in \mathbb{Z}_{\mathcal{Ν}}^{*} : x^{p}=t \right\rbrace $
\end{center}
where $t$ varies over $\mathbb{G}_{p}$. Since $ \lvert \left\lbrace x \in \mathbb{Z}_{\mathcal{Ν}}^{*} : x^{p}=t \right\rbrace \rvert \leq p$, it holds $ \lvert \mathbb{Z}_{\mathcal{Ν}}^{*} \rvert \leq  \lvert \mathbb{G}_{p}\rvert \times p$. As $\lvert \mathbb{Z}_{\mathcal{Ν}}^{*} \rvert = \mathcal{Ν}-1$, we see $p \times \lvert \mathbb{G}_{p} \rvert \geq \mathcal{Ν}-1$. Thus, in view of (11), we obtain 
\begin{equation}
\lvert \mathcal{S}_{k}^{(p)} \rvert^{2} \leq \frac{p \mathcal{Ν}^{2}}{\lvert \mathbb{G}_{p} \rvert } \leq \frac{p \mathcal{Ν}^{2}}{\frac{\mathcal{Ν}-1}{p}} \leq 2 \mathcal{Ν} p^{2}.
\end{equation}
Hence, by (12), we have proved the following

\begin{proposition} If $p$ and $q$ are natural numbers that are greater than $1$ and such that \\

\indent \hspace{5cm} $q \mid p$ or $p \mid q$,\\

\noindent then\\
\indent \hspace{3.4cm} $ \lvert \mathcal{S}_{k}^{(p)}\rvert = \lvert \sum_{x=0}^{\mathcal{Ν}-1} e^{2\pi i \frac{k x^{p} }{\mathcal{Ν}}} \rvert \leq \sqrt{2\mathcal{Ν}} p$. $\square$
\end{proposition}

\noindent
\vspace{-\baselineskip}

\end{document}